\documentclass[showpacs,twocolumn,preprintnumbers,amsmath,amssymb,prb]{revtex4}
\usepackage{graphicx}
\usepackage{dcolumn}
\usepackage{bm}
\usepackage{psfrag}
\begin{document}

\title{Monte Carlo Study of Maghemite Nanoparticles}
\author{Kenneth Adebayo}
\author{B. W. Southern}%
 \email{souther@cc.umanitoba.ca}
\affiliation{%
Department of Physics and Astronomy,\\
University of Manitoba, Winnipeg Manitoba, Canada R3T 2N2 
}%
\date{\today}
\begin{abstract}
We consider a simple model of maghemite nanoparticles and study their magnetic properties using Monte Carlo methods.
The particles have a spherical geometry with diameters ranging from 3 nm to 8 nm. The interior of the particles consists of core spins
with exchange interactions and anisotropy given by the values in the bulk material. The outer layer of the particles consists of
surface spins with weaker exchange interactions but an enhanced anisotropy. The thermal behaviour of the total, core and
surface magnetizations are calculated as well as the hysteresis loops due to the application of an applied field. The effect of the surface anisotropy
on the blocking temperature, the coercive and exchange bias fields is studied. 
\end{abstract}
\maketitle

\psfrag{b}{$\beta $}
\psfrag{n}{$\nu $}
\psfrag{D}{$\Delta $}
\psfrag{h}{$\eta $}
\psfrag{t }{$\tau $}
\psfrag{\266}{$\partial$}

\section{Introduction}
Nanomagnetism focuses on the magnetic behaviour of individual building blocks of nanostructured systems as well as on combinations 
of individual building blocks that display collective magnetic phenomena. A fundamental understanding of 
nanomagnetism will lead to the development of integrated systems with complex structures and architectures that 
possess new functionalities. Proximity effects allow multi-component composites to behave as new materials that embrace 
properties that are often mutually exclusive and thus not found in single component systems. Competing interactions 
and the presence of low-lying energetic states and quantum fluctuations help create the complexity that gives rise to 
unanticipated phenomena in magnetic nanosystems.

Magnetism and magnetic materials have been traditionally studied   with phenomenological models. These models either work well 
or must be supplemented by new terms in the model to account for unexplained effects. Such an approach has limitations at the nanoscale. 
The model may not be able to explain characteristics that depend on the details of the system at the nanoscale 
and may not have predictive capabilities. Magnetic properties at interfaces and surfaces, which make up a large 
fraction of nanostructured and confined materials, are quite different from the bulk systems upon which many simple models are built. 

Fundamental to understanding the magnetic behaviour is the evolution of the magnetism as the structural scale descends 
from the bulk to the nanoscale. Due to reduced symmetry, the magnetic anisotropy can be orders of magnitude larger than 
in the bulk. This result can lead to magnetic frustration and reorientation of the magnetization at the surface and interface. 
Understanding the complex atomic spin structure of magnetic nanostructures using computational approaches is thus 
essential to the mastering of nanomagnetism itself. Recently, magnetic measurements \cite{kas} on a collection of iron-oxide nanoparticles arranged on a macroscopic three-dimensional fcc array have been observed to exhibit a low coercivity that is weakly temperature dependent with no superparamagnetism up to 400~K, whereas only dynamical freezing and superparamagnetism is observed for a randomly packed configuration of the same  nanoparticles.  A numerical study \cite{Plumer.2009} of  the effects of  anisotropy, magnetostatic interactions and temperature on hysteresis loops for a collection of magnetic dipoles on a  fcc lattice using finite temperature micromagnetic simulations found that,  although general loop shapes resulting from the model are in agreement with these experimental results, some crucial features are not reproduced, especially as a function of temperature. The intrinsic temperature dependence of the core and surface magnetizations of individual nanoparticles needs to be incorporated to correctly describe the experimentally observed reversal processes in ordered arrays of nanoparticles. In
this work we only consider the magnetic properties of single nanoparticles in order to determine which properties might be important to understand ordered arrays of nanoparticles.

\section{Experimental Facts}

The magnetic properties of nanoparticles are greatly influenced by surface effects due to their finite size\cite{Coey,Millan} and the relative importance of surface sites increases as the size of the particle decreases. With decreasing size, surface effects lead to a decrease of the ordering temperature and an enhanced surface anisotropy which differs significantly from that of the particle core.  Recent experiments\cite{Vanlierop} on dilute suspensions of mono-disperse $\gamma$-Fe$_2$O$_3$ nanoparticles with an average diameter of 7 $nm$ have suggested that the particles block at a temperature $T_B \sim 20$~K which is much lower than that where the core develops ferrimagnetic order. Below $T_B$ the particles exhibit enhanced coercivity and exchange bias.

With decreasing size, the average magnetic coordination number of a $\gamma$-Fe$_2$O$_3$ crystallite is significantly reduced,
and finite-size effects become significant. A single-domain ferrimagnetic structure exists in the core, and the large number of Fe
ions in the unit cell make this system sensitive to vacancies distributed throughout the octahedral
sites and especially at the surface. This leads to a magnetic "roughness" which can be thought of as an effective
surface magnetocrystalline anisotropy that is different from the magnetocrystalline anisotropy of the nanoparticle single domain
core. Hence the single-domain ferrimagnetic core of the $\gamma$-Fe$_2$O$_3$ nanoparticle is magnetically
ordered at a much higher temperature than the surface spins. 

\begin{figure}[ht]
 \centering
\includegraphics[width=3.in,angle=0]{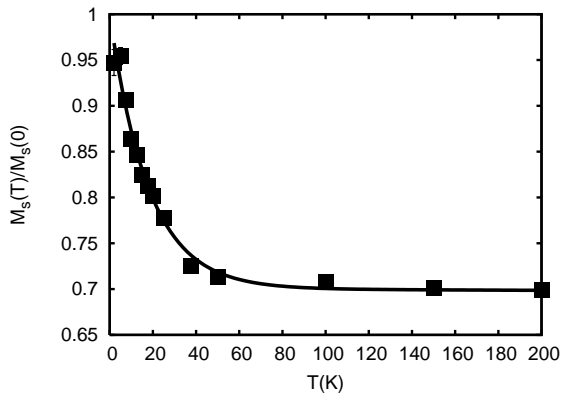}
\caption{Saturation magnetization, $M_s(T)/M_s(0)$, of $\gamma$-Fe$_2$O$_3$ nanoparticles as a function of temperature\cite{Vanlierop}.
}
 \label{fig1}
\end{figure}

A prominent experimental feature of the $7$ nm $\gamma$-Fe$_2$O$_3$ particle nanomagnetism is the unique temperature dependence of the
saturation magnetization, $M_s$, shown in Fig.~\ref{fig1}. The effect of surface spin freezing in the
$\gamma$-Fe$_2$O$_3$ nanoparticles displayed by $M_s(T )$ in Fig.~\ref{fig1}  is
well described by the phenomenological description using a modified Bloch $T^\frac{3}{2}$ law
\begin{equation}
M_s(T )/M_s(0) = (1-A)(1 - BT^\frac{3}{2}) + A e^ {-\frac{T}{T_f}}
\label{fit}
\end{equation}
which describes $M_s$ over the complete range of temperatures. Fits to the data in Fig.~\ref{fig1} with Eqn.(\ref{fit}) (shown by
the solid line) identifies a surface spin freezing temperature $T_f = 15 \pm 1$~K\cite{Vanlierop}.

\section{Model}

Bulk maghemite ($\gamma$-Fe$_2$O$_3$ ) is derived from magnetite (Fe$_3$O$_4$) which crystallizes  in an inverse spinel structure with 8 Fe$^{3+}$ ions located at tetrahedral sites ($T$), 16 Fe$^{3+}$ ions in octahedral sites ($O$), and 32 O$^{3-}$ ions per 
unit cell\cite{Brown,Costa,Dimi2,ZBoril,Dimi1}. The lattice constant of the unit cell is .83 nm. Maghemite is formed  by the removal of $\frac{1}{9}$ of the Fe atoms and  vacancies are distributed throughout the $O$ sites to ensure charge neutrality.  Hence there are $16-\frac{24}{9} = \frac{40}{3}$ Fe$^{3+}$ atoms on the $O$ sites in a unit cell. The ground state configuration of the Fe$^{3+}$ ion
is $^6S_\frac{5}{2}$ which has $S=\frac{5}{2}, L=0$ and $g_J=2$. Competing superexchange interactions between the Fe$^{3+}$ $T$ and $O$ sites leads to a ferrimagnetic ordering in the bulk material at $1020$~K. The moments on the $T$ and $O$ sites are anti-parallel and the net moment per iron atom is $( \frac{ \frac{40}{3}-8}{ \frac{40}{3}+8} )5 \mu_B =.25 g_J S \mu_B=1.25 \mu_B$.

In order to study the magnetic properties of maghemite nanoparticles, we shall use the following Hamiltonian
\begin{equation}
\small 
H=-\sum_{i<j} J_{ij} \vec{S}_i \cdot \vec{S}_j - K_s \sum_{i\in s}(\vec{S}_i 
\cdot \hat{n_i})^2 - K_c \sum_{i\in c}{S}^2_{iz} - H_z \sum_{i} S_{iz}
\label{ham} 
\end{equation}
where the first term describes the exchange interactions with the nearest magnetic neighbours of each iron site, the second 
term describes  the single-ion uniaxial surface anisotropy, the third term 
accounts for the uniaxial core anisotropy and the last term is an applied magnetic 
field. The unit vector $\hat{n_i}$ is the local normal to the surface of the particle. In the spinel structure the tetrahedral sites have 12 nearest neighbours
on the octrahedral sites and 4 nearest neighbours on tetrahedral sites whereas the octrahedral sites have 6 nearest neighbours of each type. The corresponding superexchange interactions were taken to be antiferromagnetic with the values  $J_{TT}=-42.0K, J_{TO}= -56.2K$, and $J_{OO}= -17.2K$. These values were taken from
the literature for bulk maghemite\cite{Berk,Kodama,Linder}.  Since the electronic, structural and magnetic properties of nanoparticles are modified in the surface region, we expect an enhanced anisotropy and weaker exchange.
However, the values of these parameters are not well known. For this reason, we make the reasonable assumption that the surface-core  exchange interactions are all weaker by factor of two  and the surface-surface interactions are weaker by a factor of ten. The single-ion site surface anisotropy $K_s$ was given the values $1.0 ,5.0$ and $10.0$ ,while  the core anisotropy $K_c = 0.02$. The spins at each site are taken to be classical spins of unit magnitude and vacancies on the octrahedral sites
are given a spin magnitude of zero. The oxygen ions are considered to be non-magnetic and only serve to provide superexchange interaction pathways between
the occupied Fe sites. The actual spin of each iron atom is $S=\frac{5}{2}$ and is large enough to be treated classically at finite temperatures. However,
important quantum effects are present at low temperatures. Since our model uses classical unit vectors for the $Fe$ ions whereas the actual spin length is $J=g_J S=5$, there is an extra factor of $J(J+1)/3=10$ in the temperature scale that has not
been included in the definition of our exchange constants.

 \begin{figure}[ht]
 \centering
\includegraphics[width=4.in,angle=0]{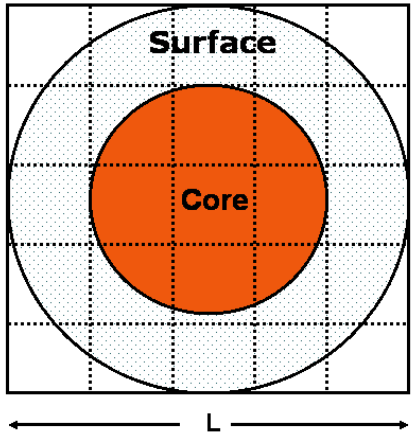}
\caption{Geometry of nanoparticles}
 \label{fig2}
\end{figure}

We take the maghemite particles to have a spherical geometry with a diameter of $L$ unit cells as shown in Fig.~\ref{fig2}. 
The $Fe^{3+}$ ions in the outermost shaded cells are called surface sites and the ions in central cells are core sites. For $L=5$ the diameter is about $4.2 \  nm$ and the particle contains 1112 surface sites, 311 core sites and 170 vacancies. For this size of particle, the majority of  sites are on the surface.

\begin{table}[htbp]
\centering 
\caption{Number of Surface, Core and Vacant Sites} 
\begin{tabular}{cccccc} \hline 
$L$ & $Diameter(nm)$ & $Surface $ & $Core $ & $Total $ & $Vacancies$\\ \hline 
4 &3.3 &667 &94 &761&80 \\
5 &4.2 &1112 &311 &1423&170 \\ 
6 &5.0 &1683 &751 &2434&297 \\ 
7 &5.8 &2467 &1411 &3878&473 \\ 
8 &6.6 &3358 &2440 &5798&687 \\ 
9 &7.5 &4327 &3891 &8218&971 \\ \hline 
\end{tabular}
\label{sites} 
\end{table}

We study particles  with sizes ranging from $L=4$ to $L=9$ (3 to 8 $nm$ diameter). 
The number of surface, core and vacant sites for $L=4$ to $9$ are given in Table \ref{sites}. The vacancies are distributed
uniformly throughout the octahedral sites in both the core and surface and account for approximately $1/9$ of the total sites. 
From  Table \ref{sites} we see that the majority of the sites are on the surface. For a perfect ferrimagnetic order where the occupied tetrahedral sites have their spins anti-parallel to those on the octrahedral sites, we can use the data in Table \ref{sites} to predict the values of the  core, surface and total magnetization per core, surface and total number of sites respectively as shown in Table \ref{mags}.

\begin{table}[htbp] 
\centering 
\caption{Predicted Core, Surface and Total Magnetizations} 
\begin{tabular}{cccc} \hline 
$L$ & $m_{core}$ & $m_{surface}$ & $m_{total}$ \\ \hline 
4 &.255&.262&.261 \\
5 &.209&.277&.262 \\ 
6 &.252&.247&.248 \\ 
7 &.256&.275&.268 \\ 
8 &.250&.265&.259 \\ 
9 &.271&.241&.255 \\ \hline 
\end{tabular}
\label{mags} 
\end{table}  

\section{Monte Carlo Results}

We will use Monte Carlo techniques to study the magnetic properties of these nanoparticles. There have been many previous theoretical studies of
the magnetic properties of nanoparticles. The group of Restrepo et. al. \cite{Labaye1,Labaye2,Labaye3,Labaye4} have used the Metropolis
algorithm \cite{metrop} to study the effect of surface anisotropy on the sublattice magnetizations with uniform exchange interactions throughout
the particles. Trohidou et. al. \cite{Eftax,Troh,Vasi} have also used the Metropolis algorithm to study nanoparticles with a ferromagnetic core surrounded by an antiferromagnetic
surface layer.
Iglesias et. al. \cite{Oscar1,Oscar2,Amilcar,Igle} have studied both systems with a ferromagnetic core surrounded by an anti-ferromagnetic surface layer as well as a more realistic model of maghemite with the inverse spinel structure. Several other groups \cite{Kach,biasi,Mazo,Mazo1,Mazo2,Usov} have
also used the conventional Metropolis algorithm  to study the effects of surface anisotropy on the magnetic properties of maghemite nanoparticles.

We employ a heat bath algorithm that is a modified version of the usual heat bath method used to study classical spins to allow
for the finite length of the $Fe$ spins in maghemite. The modifications only affect the behaviour of the thermodynamic quantities at low
temperatures where classical models fail. 
The heat bath Monte Carlo method\cite{Miyatake,Lee} generates a sequence of configurations which simulate a canonical ensemble of states in thermal equilibrium at a constant temperature $T$. Given a configuration of the spins in a system, the direction of each spin  is updated assuming that the spin is in  contact with a heat bath that puts it into local equilibrium with the surrounding spins. At each Monte Carlo step, there is a spin update and hence no moves are rejected. This is a very efficient way to sample the accessible microstates.  In the case of classical spins, the spin direction with respect  to the local field $\vec{H}_{i}^{eff}$ can be described using the azimuthal and polar angles   $\phi_{i}$ and $\theta_{i}$ respectively.
The heat bath algorithm usually allows the cone angle $\theta$ to have any value in the range $[0,\pi]$. As shown below in Fig.~\ref{fig3}, a semi-classical picture of the possible spin orientations in a field is a set of cones with the local field along the cone axis

\begin{figure}[!hpb]
 \centering
\includegraphics[width=3.5in,angle=0]{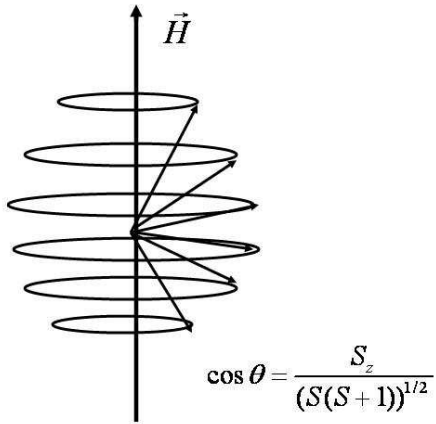}
\caption{Semiclassical description of spin with $S=5/2$}
\label{fig3}
\end{figure}

We use the
heat bath algorithm to generate a value of $\cos \theta$ but we then place the spin direction relative to the effective field onto the nearest cone.
This procedure has little effect at high temperatures but does modify the low temperature dependence of the magnetization and energy.
We start the simulation at high temperature and cool in either a zero or applied field. We measure the energy, core and surface magnetizations as a function of temperature. At low temperatures we also measure hysteresis loops. Typically we use $30,000 \ mcs$ for our measurements at each value of $H$ and $T$. We have studied particles with sizes ranging from $L=4$ to $9$ (3 to 8 $nm$ diameter).

\subsection{Thermal Dependence of the Magnetization}
The temperature dependence of the magnetization can give important information about the properties of the nanoparticles.
In the classical model we have a unit vector at each Fe site in the nanoparticle. The core, surface and total magnetizations are given by
\begin{eqnarray}
\vec{m}_{core}&=&\frac{1}{n_{core}} \sum_{i \in core} \vec{S}_i \nonumber \\
\vec{m}_{surf}&=&\frac{1}{n_{surf}} \sum_{i \in surf} \vec{S}_i  \\
\vec{m}_{tot}&=&\frac{1}{n_{tot}} \sum_{i \in tot} \vec{S}_i  \nonumber
\end{eqnarray}
where $n_{core}, n_{surf}$ and $n_{tot}$  are the number of occupied core, surface and total sites as given in Table \ref{sites}. We have calculated average values
of each component of the magnetization as well as the magnitude squared at each temperature. The particle is initialized
at high temperature with the spins in random directions. The Monte Carlo updates are then performed for $1000$ Monte Carlo steps (mcs)
and the measurements are collected over the next $5000$ mcs at this temperature. The temperature is then lowered and the process is repeated.

\begin{figure}[htbp]
 \centering
\includegraphics[width=2.5in,angle=-90]{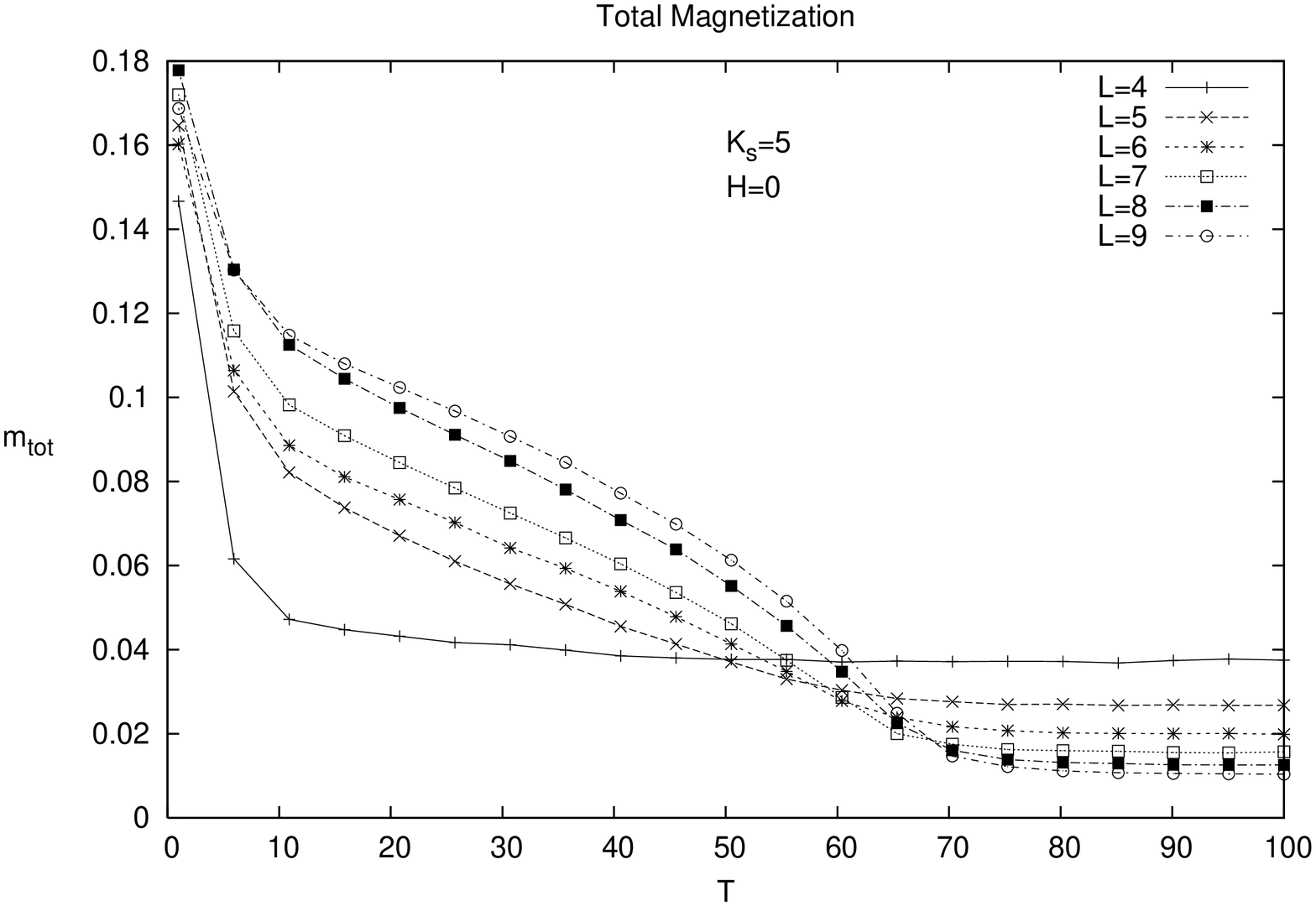}
\includegraphics[width=2.5in,angle=-90]{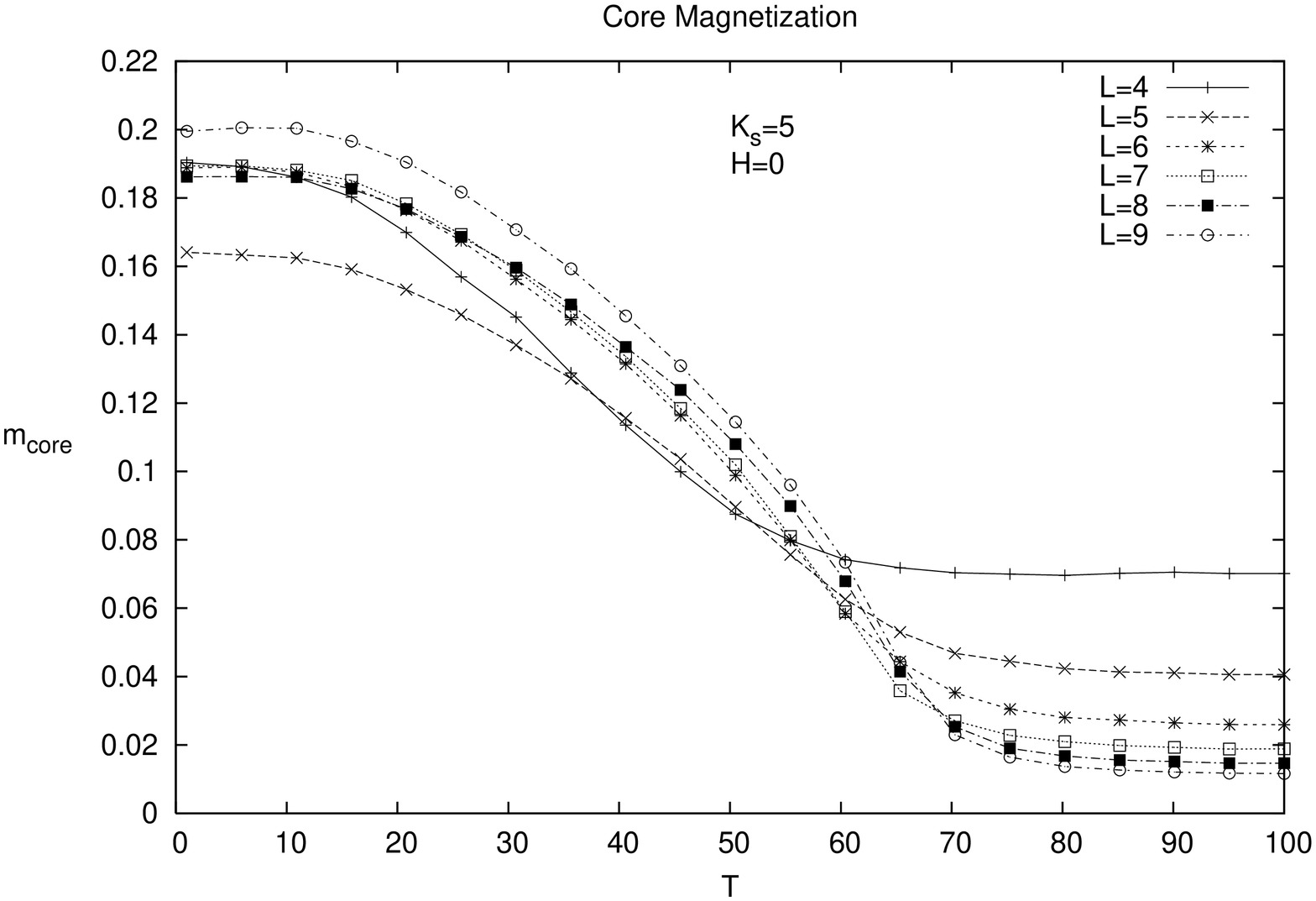}
\includegraphics[width=2.5in,angle=-90]{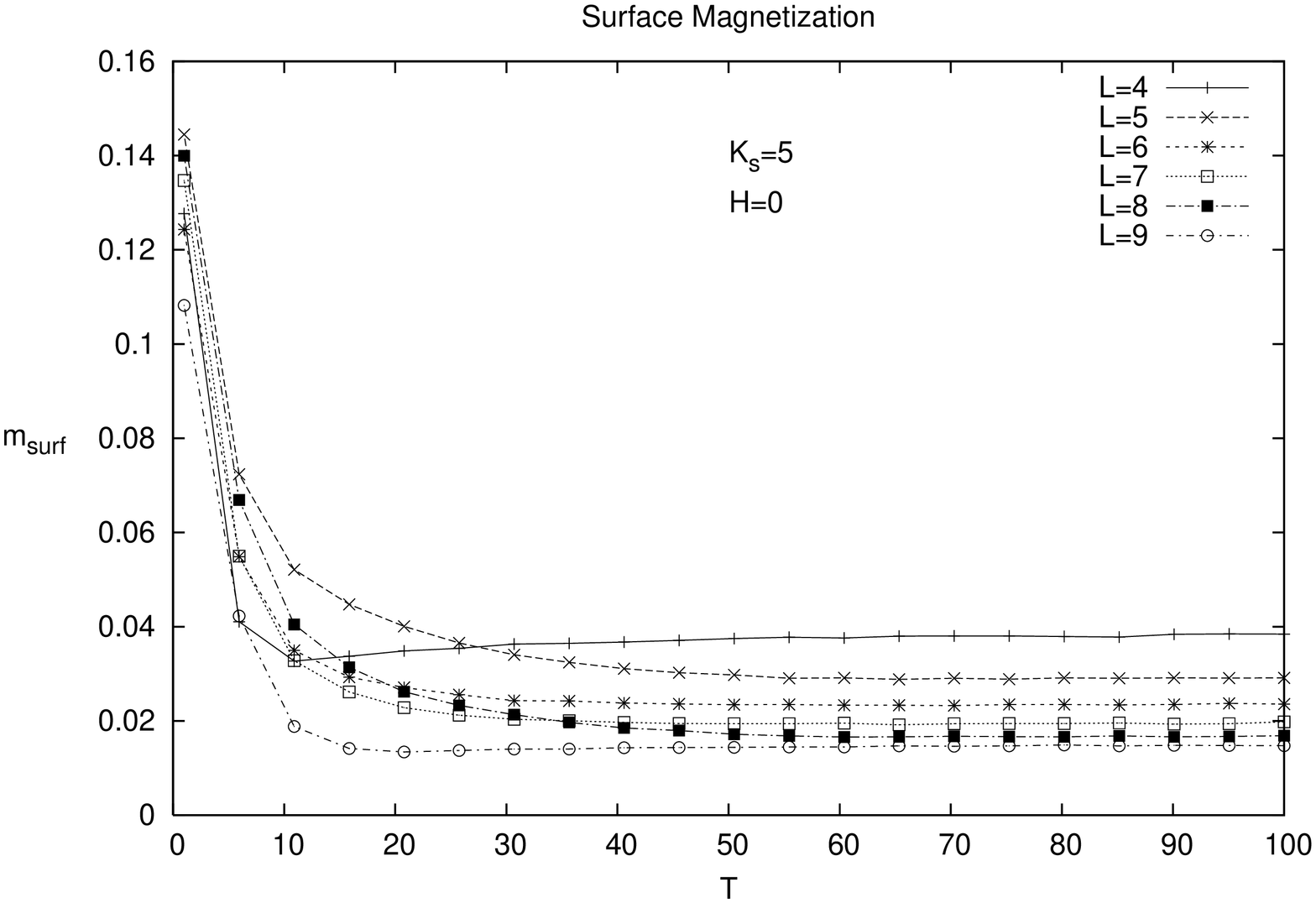}
\caption{Magnitudes of the total magnetization (upper panel), the core (middle panel) and the surface (lower panel) magnetizations as a function of temperature for particle sizes ranging from $L=4$ to $L=9$.
The surface anisotropy constant has the value $K_s=5$ and the applied field $H=0$.
}
 \label{fig4}
\end{figure}


The upper panel of Fig. \ref{fig4} shows a plot of the magnitude of the total magnetization for particle sizes ranging from $L=4$ to $L=9$ as a function of $T$ in the zero field cooled case. The surface
anisotropy constant has the value $K_s=5$. At high $T$, the magnetization magnitudes do
not vanish but rather approach $\frac{1}{\sqrt{n_{tot}}}$ with the larger values of $L$ lying below those of smaller $L$. This is
a finite size effect.  At lower temperatures, the magnetizations
with smaller $L$ lie below those of larger $L$. The crossover  for a given pair of sizes occurs at temperatures in the range from $T=50$ to $T=70$. For the largest size pair, the crossover is near $T_c \sim 68$. We identify this temperature with the ordering
temperature of the particle where strong correlations between neighbouring spin directions develop. This value of $T_c$ differs from the experimental value of $T_c$ for maghemite particles of these sizes by an order of magnitude because our model uses classical unit vectors for the $Fe$ ions whereas the actual spin length is $J=g_J S=5$. As mentioned previously, including this factor would yield an ordering temperature of the order of $680$~K which does lie in the range of the experimental values. 

 Below $T_c$, the total magnetization
increases with decreasing temperature and displays an even more rapid increase below about $T\sim 5$. 
The total magnetization includes contributions from both the core and surface sites. Although the core develops order at $T_c$, the surface sites
can remain disordered until much lower temperatures. The Monte Carlo approach allows us to separate these two contributions. The middle and lower panels of Fig. \ref{fig4} show
the magnitudes of both the core and surface magnetizations as a function of $T$.
The core magnetizations begin to develop at about $T_c \sim 68$ and saturate at low $T$. The values of the
magnetizations at $T=0$ are
slightly less than those predicted in Table \ref{mags} and indicate we have an imperfect ferrimagnetic order
in the core. In contrast,
the surface magnetizations remain zero down to very low temperatures and only  increase substantially only below $T \sim 5$. The particles with small $L$ are dominated by surface
sites and only have a small number of core sites. The particles have ferrimagnetic order with a small amount of disorder
due to the oxygen vacancies on the octrahedral sublattice which weaken the exchange interactions. The surface sites are also affected by the reduced coordination number at the spherical surface. Note that the  $T=0$ core magnetizations are
reduced much more for small $L$ and that the surface magnetizations are reduced even more than those of the core for all sizes.

\begin{figure}[htbp]
 \centering
\includegraphics[width=2.5in,angle=-90]{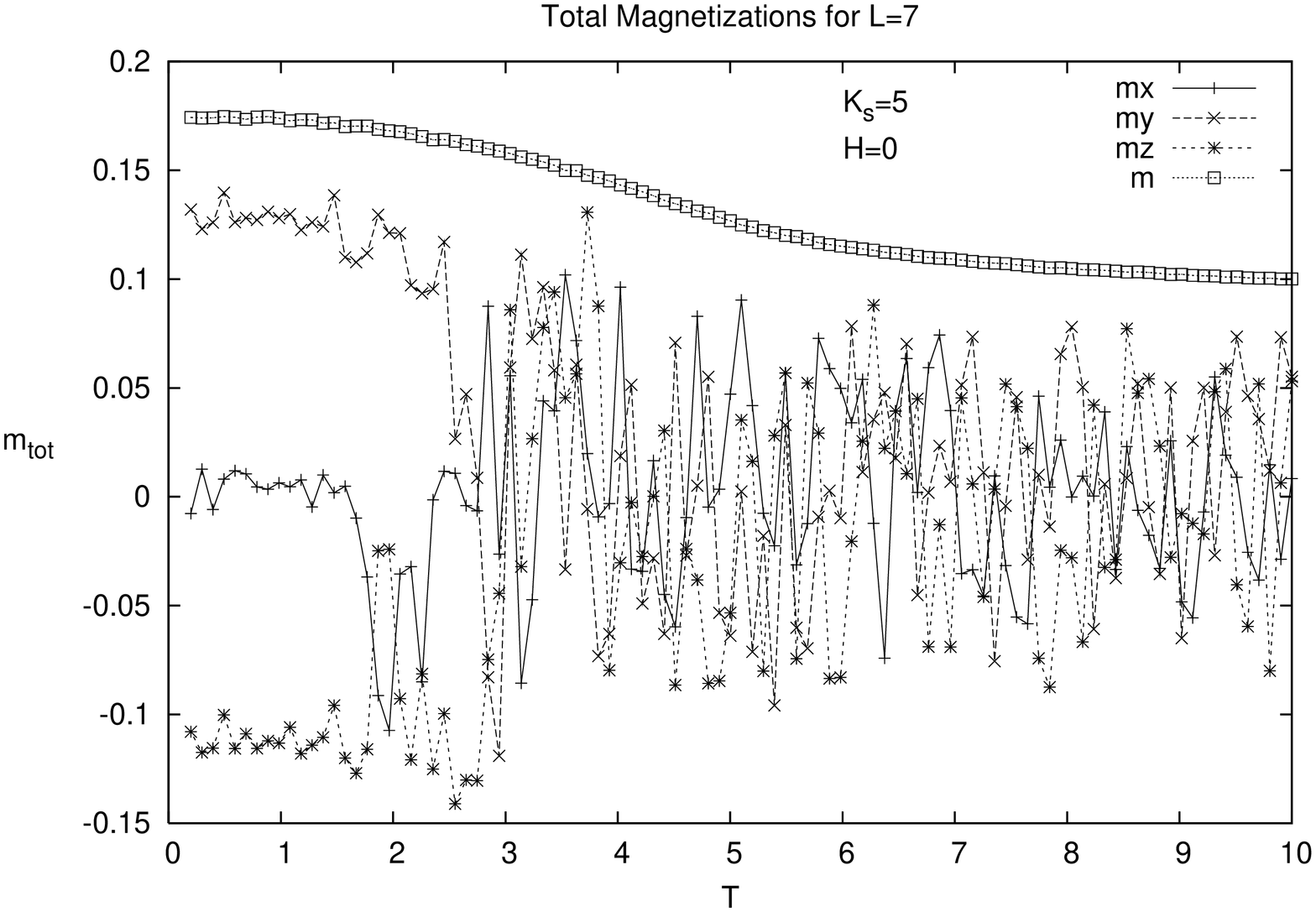}
\includegraphics[width=2.5in,angle=-90]{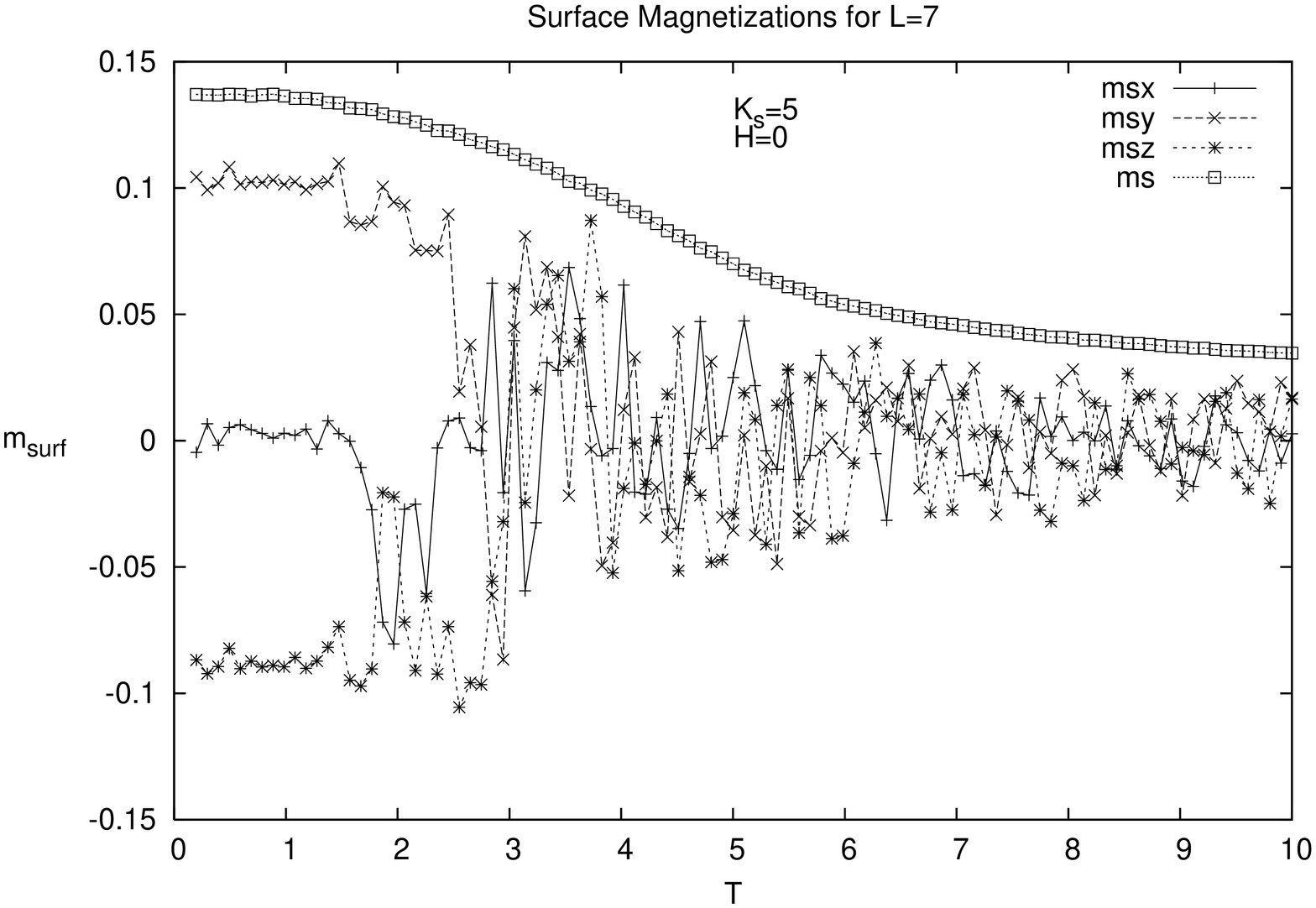}
\includegraphics[width=2.5in,angle=-90]{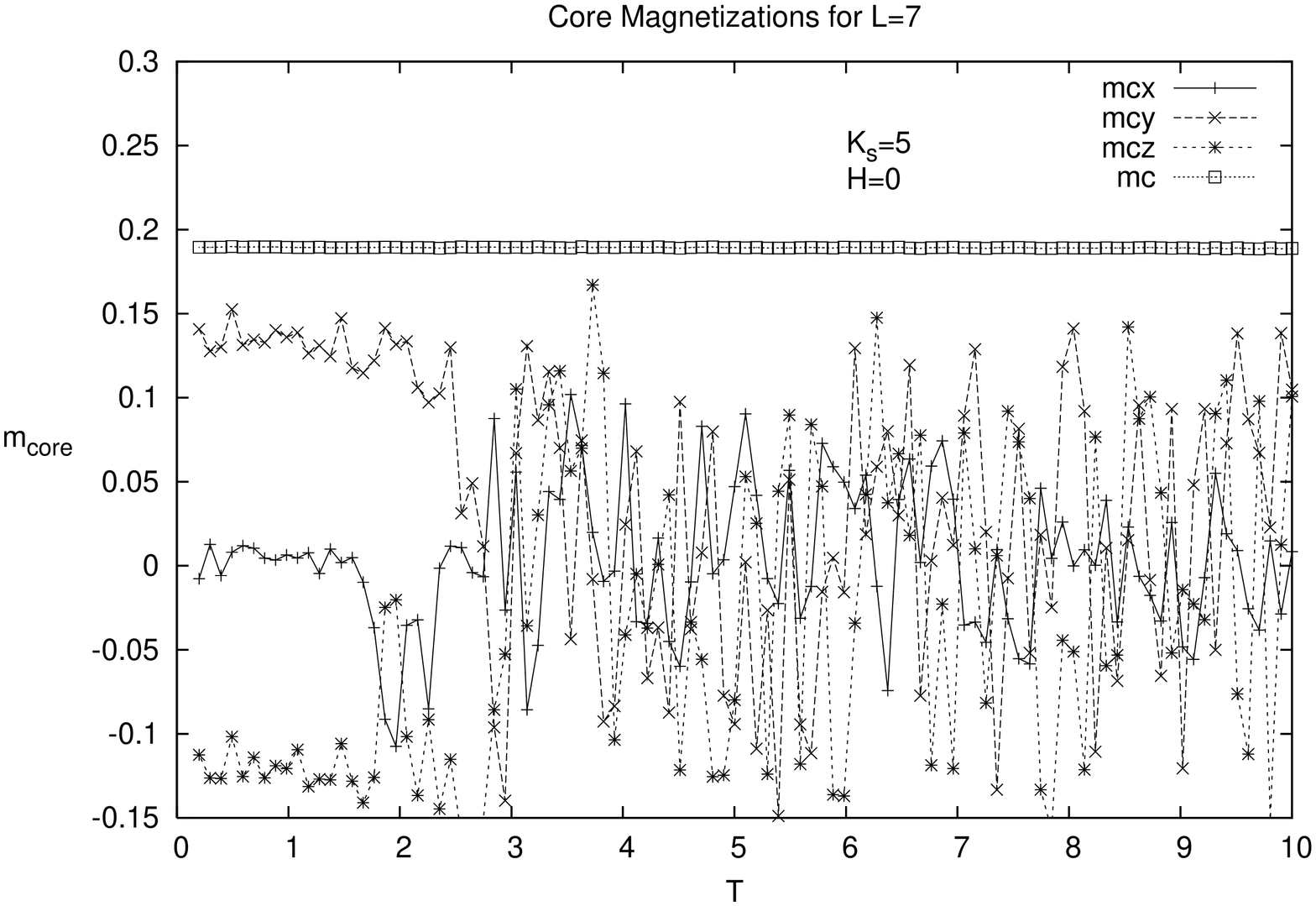}
\caption{Components and magnitude of the total (upper panel), surface (middle panel) and the core (lower panel) magnetization as a function of temperature for particle size $L=7$.
The surface anisotropy constant has the value $K_s=5$ and the field $H=0$.
}
 \label{fig5}
\end{figure}

Additional information can be obtained by measuring the vector components of the total magnetization as the temperature is reduced. Fig. \ref{fig5} shows the 
components and the magnitude of the total magnetization  as a function of $T$ for $L=7$ at low temperatures. At higher
temperatures $T > 70$,
all the components average to zero  which indicates that the particles are paramagnetic in this temperature range.  Below $T_c \sim 68$, the components of $\vec{m}_{tot}$ fluctuate from
one temperature to another with an increasing amplitude and the magnitude increases smoothly. This indicates that the particle is developing ferrimagnetic order but it is not blocked in this range of temperature. Although the
particle has a net moment, it is superparamagnetic since the direction of the net moment can overcome thermal
barriers and change its direction spontaneously. At much lower temperatures (upper panel), the components of $\vec{m}_{tot}$ cease to fluctuate and
become blocked below the blocking temperature $T_B \sim 3$.

The middle and lower panels of Fig.~\ref{fig5} show the components and magnitudes of the surface and core magnetizations at low $T$ for $L=7$. In both cases, the magnitudes vary smoothly but the components fluctuate from one temperature to another until below $T_B \sim 3$ where both the surface and core magnetizations become blocked
and cannot overcome the thermal barriers. The surface magnetization appears to develop at a slightly higher temperature $T_s \sim 6$ which is
about $10\%$ of $T_c$. All of these results are for a value of the surface anisotropy $K_s=5$.  Similar behaviour is observed for
all particle sizes.

We have repeated the calculations with a smaller values of the surface anisotropy constant. For $K_s=1$, the components of the magnetizations exhibit strong fluctuations down to much lower temperatures than for $K_s=5$.  The surface ordering temperature is reduced to $T_s\sim 3$  and the blocking temperatures are reduced to $T_B \sim 0.5$.
For larger values of the surface anisotropy, the surface ordering and blocking temperatures are increased.  For $K_s=10$, the blocking temperatures are increased to $T_B \sim 4.5$  and the surface appears to order at $T_s\sim 8$. This behaviour is again characteristic of all particle sizes. For all values of $K_s$ and $L$, the core and surface
magnetizations appear to block at the same temperature but have different ordering temperatures.

\begin{table}[htbp] 
\centering 
\caption{Blocking Temperature $T_B$ for various $L$ and $K_s$} 
\begin{tabular}{cccc} \hline 
$L$ & $K_s=1$ & $K_s=5$ & $K_s=10$ \\ \hline 
4 &0.3 &2.9&4.0 \\
5 &0.4&3.2&4.2 \\ 
6 &0.4&3.0&4.4 \\ 
7 &0.5&3.0&4.5 \\ 
8 &0.5&2.7&4.5 \\ 
9 &0.5&3.0&4.5 \\ \hline 
\end{tabular}
\label{TB} 
\end{table}

The values of $T_B$ estimated from the temperature dependence of the magnetization components are tabulated in Table \ref{TB} for the different values of $L$ and $K_s$. The blocking temperature seems to be fairly independent of $L$, which is a measure of the particle size, but increases with $K_s$. The blocking temperature of a single domain particle with an effective anisotropy $K_{eff}$ and volume $V$ is expected to be proportional\cite{neel,brown5} to $K_{eff}V$ . The effective anisotropy is usually assumed
to be both temperature and size independent and hence one might expect $T_B$ to increase with particle size. However,
several studies on nanoparticles \cite{bodker,skoropata,he} indicate that $K_{eff}$ increases as the particle size decreases. The decrease in $K_{eff}$ with size can compensate for the expected increase with $V$ and lead to a size-independent value of $T_B$. A decrease in $K_{eff}$ with $V$ could also be due to stronger temperature renormalization in smaller particles. Our results indicate that the effective energy barrier is independent of
the particle size over the range of $L$ studied.

The surface magnetizations for $K_s=5$ shown in the lower panel of Fig.~\ref{fig4} can be fit to a expression of the form
\begin{equation}
m_{surf}(T)=a \exp(-T/T_s) +b
\end{equation}
where $b$ describes the finite size effects at high $T$ and for each $L$ has the value $\frac{1}{\sqrt{n_{surf}}}$. The fits were carried out over the temperature
range $[1,50]$ and the fitting parameters are tabulated in Table \ref{TS}. The values of $T_s$ are slightly larger than the
values of $T_B$ determined from the components of the magnetization and indicate the temperature where the surface
spins develop order but are not yet blocked.

\begin{table}[htbp] 
\centering 
\caption{Fit  to $m_{surf}(T)=a e^{-T/T_s}+b$ \  for $K_s=5$.} 
\begin{tabular}{cccc} \hline 
$L$ & $a$ & $b$ & $T_s$ \\ \hline 
4 &0.13 $\pm$ 0.01&0.034 $\pm$ 0.094&2.91 $\pm$ 0.39 \\
5 &0.13 $\pm$ 0.01& 0.033 $\pm$ 0.001&5.83 $\pm$ 0.40 \\ 
6 &0.13 $\pm$ 0.01&0.024 $\pm$ 0.001&4.75 $\pm$ 0.35 \\ 
7 &0.15 $\pm$ 0.01&0.020 $\pm$ 0.001&4.69 $\pm$ 0.32\\ 
8 &0.15 $\pm$ 0.01&0.019 $\pm$ 0.001&5.93 $\pm$ 0.35 \\ 
9 &0.13 $\pm$ 0.01&0.013 $\pm$ 0.001&4.29 $\pm$ 0.36 \\ \hline 
\end{tabular}
\label{TS} 
\end{table}

However, if we increase the number of temperature points in the range $ [0.2,10] $, then
the exponential form is not appropriate. The modified form

\begin{equation}
m_{surf}(T)=a \exp(-(T/T_s)^n) +b
\end{equation}

\noindent works extremely well. Table \ref{TS1} shows the values of $T_s$ and the exponent $n$ for $K_s=1, 5$ and $10$.
The value of $T_s$ increases with $K_s$ in much the same way as $T_B$ but is fairly independent of $L$. $T_s$
is larger than $T_B$ which seems to indicate that the surface develops order before the particle becomes blocked as
the temperature is reduced. The exponent $n$ is also independent of $L$ but increases with the value of $K_s$.

\begin{widetext}
\begin{center}
\begin{table}[htbp]
\centering 
\caption{Fit  to $m_{surf}(T)=a e^{-(\frac{T}{T_s})^n}+b$ for $K_s=1,\ 5, 10$} 
\begin{tabular}{ccccccc} \hline
\multicolumn{1}{c} {\ } &
\multicolumn{2}{c} {$K_s=1$} & 
\multicolumn{2}{c} {$K_s=5$} &  
\multicolumn{2}{c} {$K_s=10$}   \\ \hline 
$L$ & $T_s$ & $n$ & $T_s$&$n$&$T_s$&$n$ \\ \hline 
4 &2.92 $\pm$ 0.01&2.30 $\pm$ 0.02&4.40 $\pm$ 0.01 &3.02 $\pm$ 0.04 & 5.99 $\pm$ 0.02 & 3.75 $\pm$ 0.05\\
5 &3.25 $\pm$ 0.03& 1.84 $\pm$ 0.04&4.82 $\pm$ 0.03 &2.60 $\pm$ 0.05 & 6.19 $\pm$ 0.03 & 3.32 $\pm$ 0.05\\ 
6 &3.38 $\pm$ 0.02&1.96 $\pm$ 0.04&4.99 $\pm$ 0.02 &2.86 $\pm$ 0.04&6.34 $\pm$ 0.02&3.72 $\pm$ 0.04\\ 
7 &3.34 $\pm$ 0.02&1.88 $\pm$ 0.03&4.88 $\pm$ 0.02 &2.64 $\pm$ 0.03&6.24 $\pm$ 0.01&3.27 $\pm$ 0.02\\ 
8 &3.64 $\pm$ 0.02&1.81 $\pm$ 0.03&5.19 $\pm$ 0.02 &2.54 $\pm$ 0.03&6.57 $\pm$ 0.01& 3.19 $\pm$ 0.02\\ 
9 &3.59 $\pm$ 0.02&1.92 $\pm$ 0.03&5.15 $\pm$ 0.01 &2.69 $\pm$ 0.02&6.54 $\pm$ 0.01& 3.27 $\pm$ 0.02\\ \hline 
\end{tabular}
\label{TS1} 
\end{table}
\end{center}
\end{widetext}

Both $T_B$ and $T_s$
display the same qualitative behaviour with $T_s > T_B$ as a function of $L$ and $K_s$. This suggests that the particle
blocking is associated with the surface ordering and that the surface anisotropy plays an important role in both quantities.

\subsection{Field Cooling}

\begin{figure}[htbp]
 \centering
\includegraphics[width=2.5in,angle=-90]{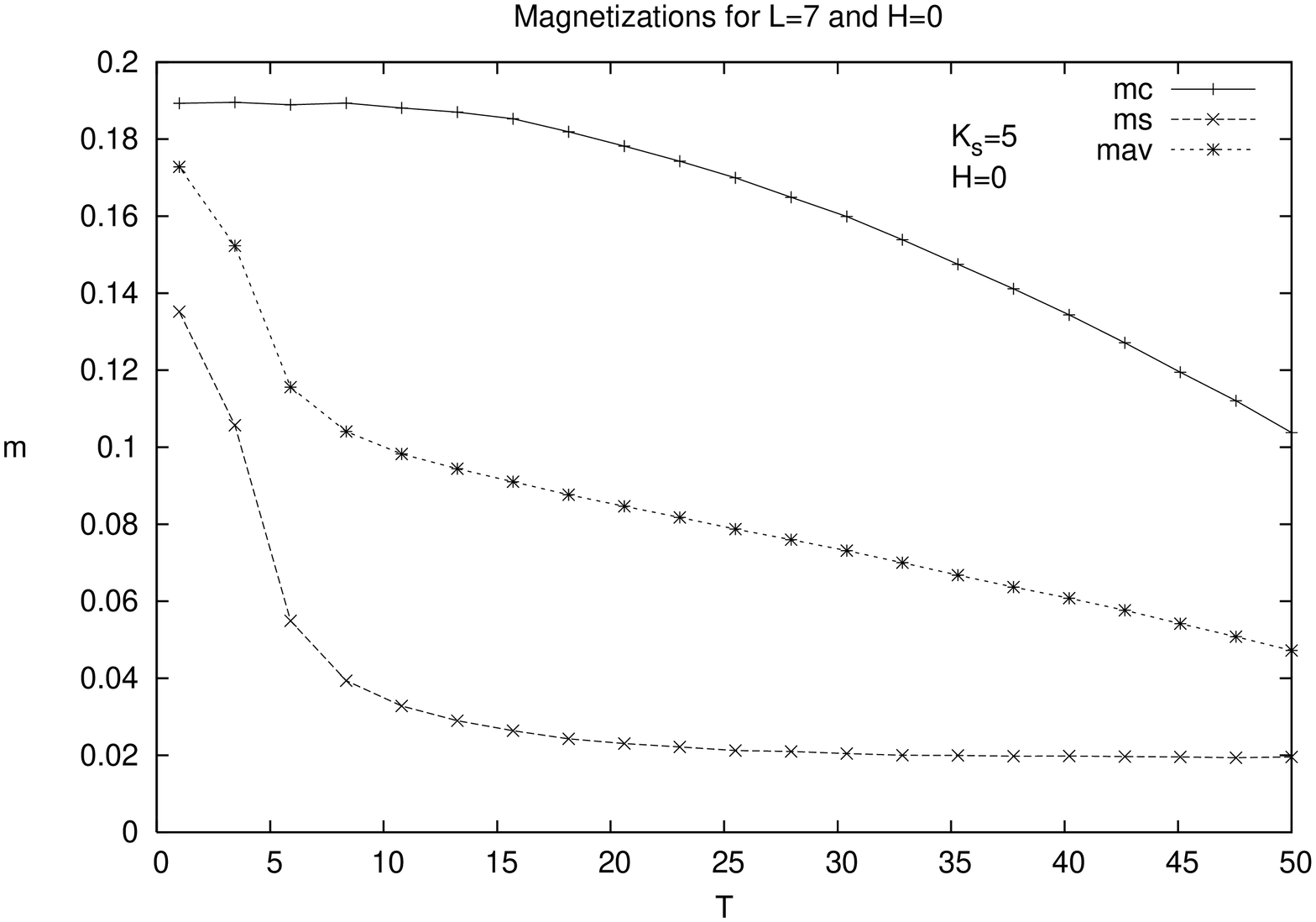}
\includegraphics[width=2.5in,angle=-90]{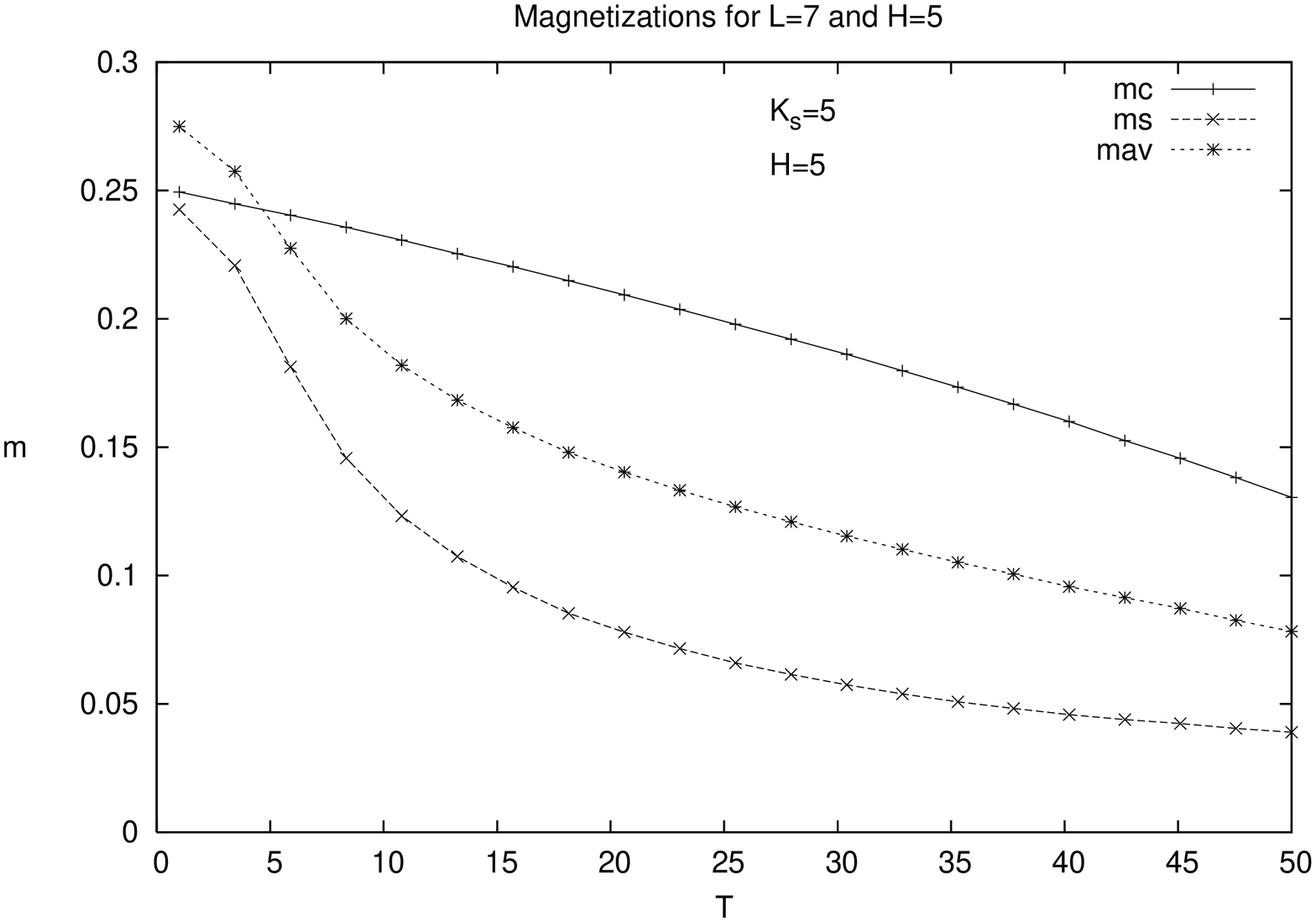}
\caption{Magnitudes of the total, core and surface magnetizations as a function of temperature for particle size $L=7$ when cooled in a
field of magnitude $H=0$ (upper panel) and a field $H=5$ (lower panel).
The surface anisotropy constant has the value $K_s=5$ in both cases.
}
 \label{fig6}
\end{figure}

We have also studied the magnetic properties of the nanoparticles when they are cooled in the presence of an applied field.
The field cooling procedure starts from a random spin configuration at high $T$ and then the temperature is slowly reduced to $T=1$. During the
cooling, we measure the total, core and surface magnetizations. Fig. \ref{fig6} shows the magnetizations for both the zero field and non-zero field
cases with $L=7$ and $K_s=5$. The lower panel of the figure shows that  the magnetizations are  increased in the field cooled case.
At the lowest temperature, the surface magnetization has almost doubled compared to the zero field case. The magnitude of the applied field
is the same as the magnitude of the surface anisotropy  and these contributions to the energy of the particle are in competition.

\begin{figure}[htbp]
 \centering
\includegraphics[width=2.5in,angle=0]{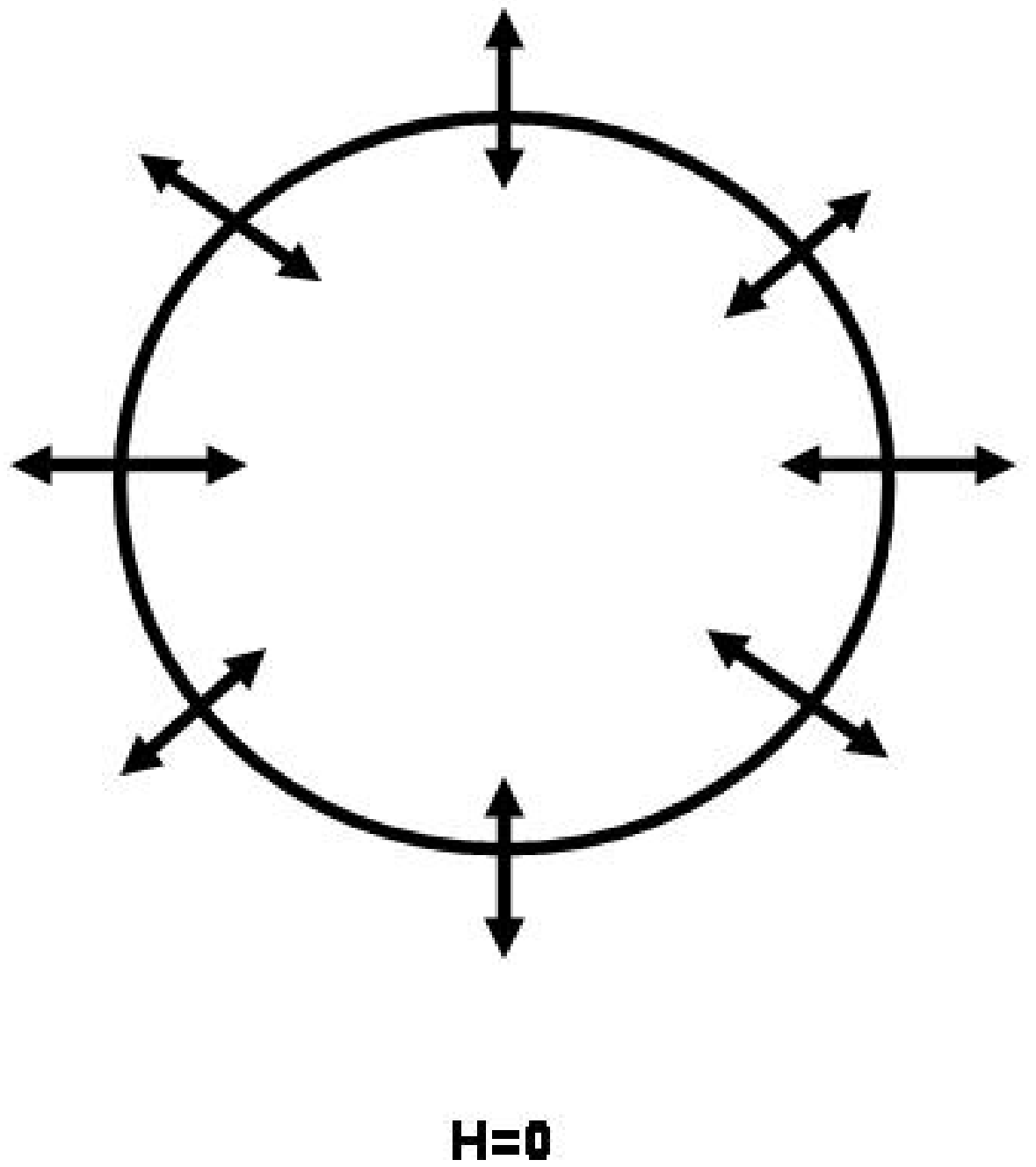}
\includegraphics[width=2.5in,angle=0]{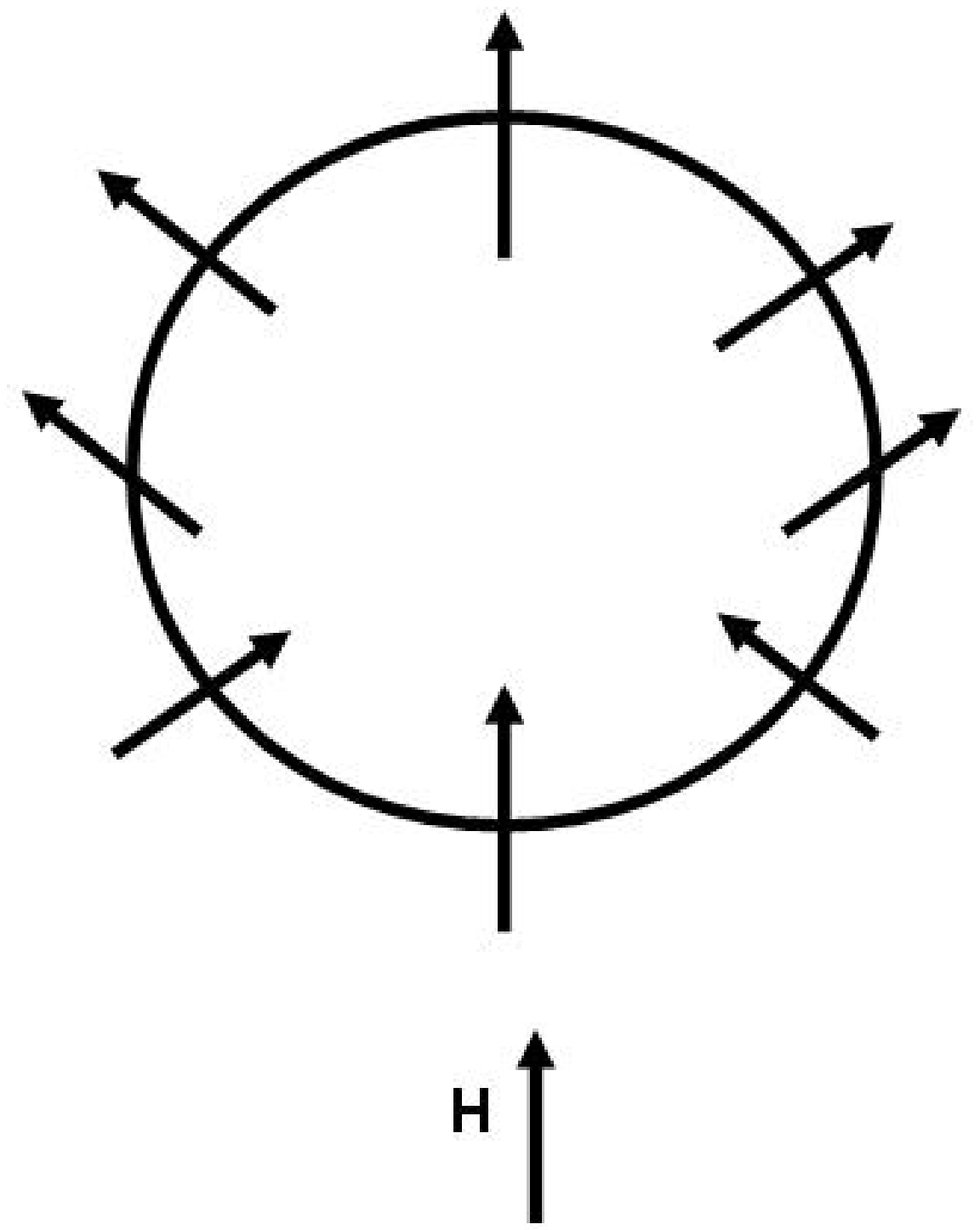}
\caption{Zero-field cooled and non-zero field cooled configurations}

 \label{fig7}
\end{figure}

The surface anisotropy favours the surface moments to be normal to the surface whereas the applied field favours alignment in a particular direction.
When cooled in zero field, the surface anisotropy leads to some of the surface moments pointing outward from the surface and some inward
as both satisfy the normal arrangement. Hence the surface anisotropy can lead to a net moment reduction. In the presence of a field,
the direction of the surface moments will change and will point outward or inward according to whether the outward normal has a component
parallel or anti-parallel to the field. The net effect is an increase in the net surface moment as shown schematically in Fig. \ref{fig7}.

\subsection{Hysteresis Loops}

After cooling the particle in either zero field or non-zero field, we have measured the hysteresis properties of a nanoparticle. In each
case we begin at the lowest temperature with a field applied in the $z$-direction and slowly reduce its magnitude to zero and then  slowly increase the field in the reversed direction until the magnitude is the same as it was at the beginning. This process is then carried out in the reverse direction.
At each value of the applied field, we perform 30,000 mcs to measure the components of the magnetization and the energy. Magnetic contributions to the total magnetization per site arising from the surface and core sites were computed separately.
 
\begin{figure}[htbp]
 \centering
\includegraphics[width=2.5in,angle=-90]{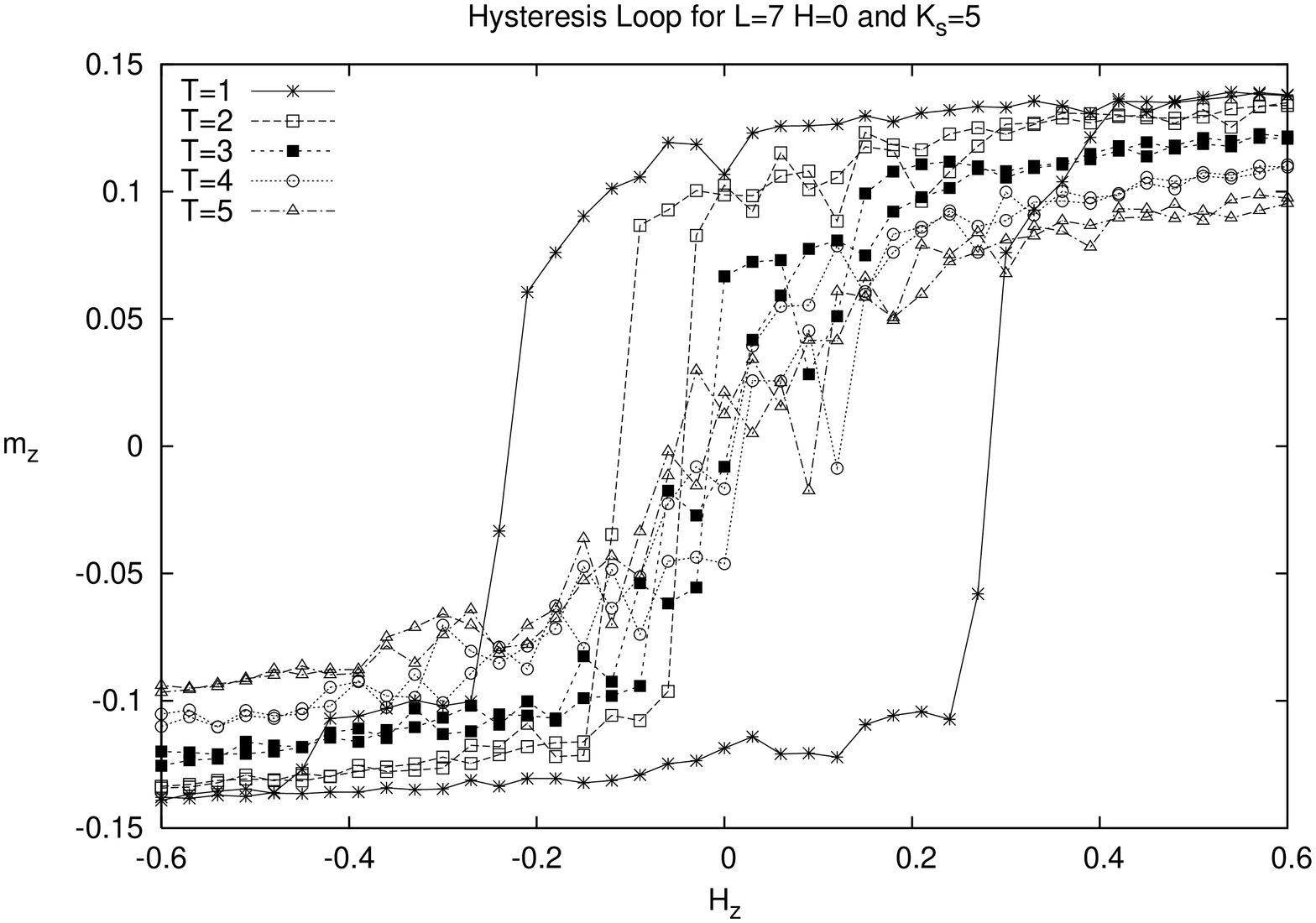}
\includegraphics[width=2.5in,angle=-90]{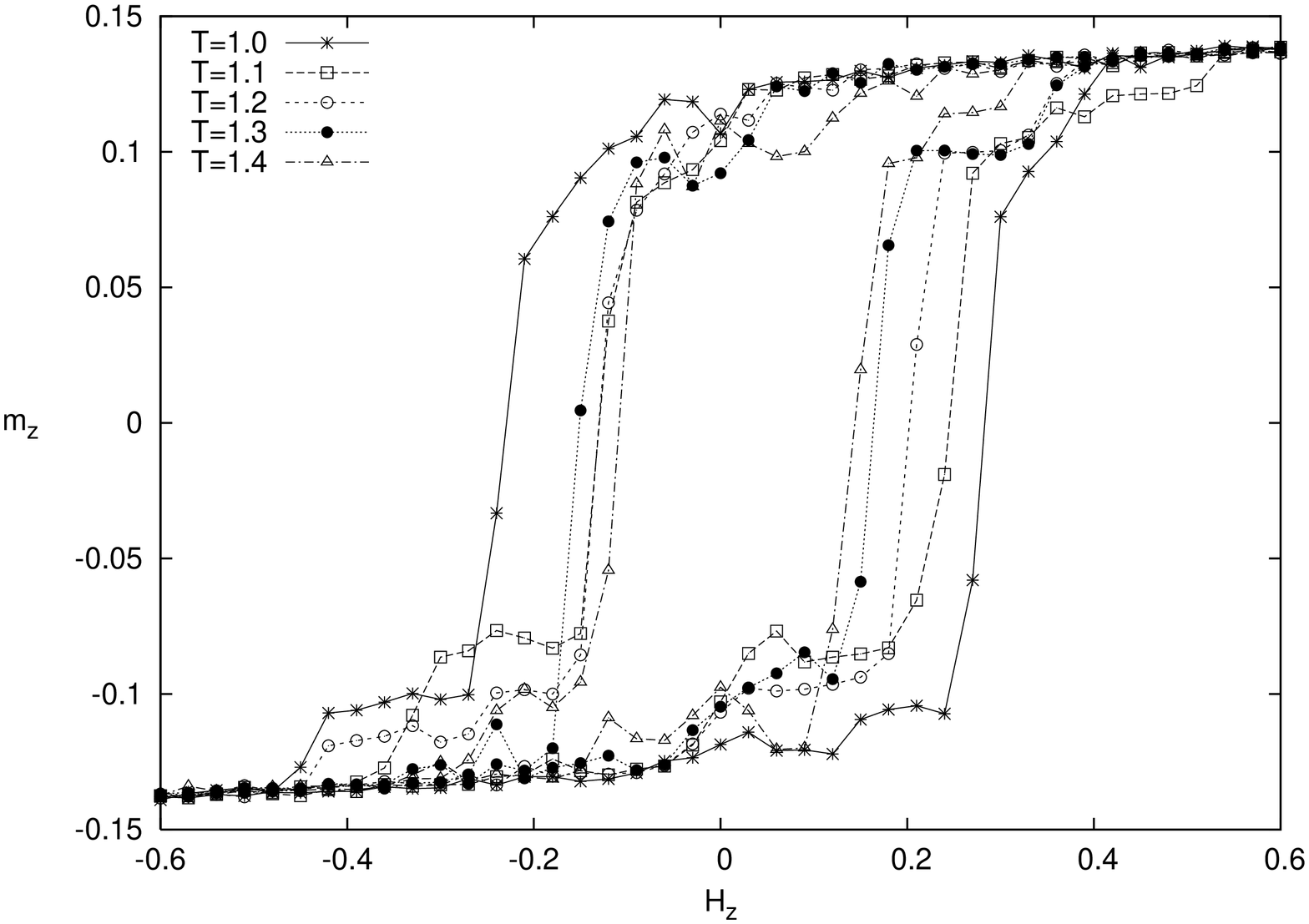}
\includegraphics[width=2.5in,angle=-90]{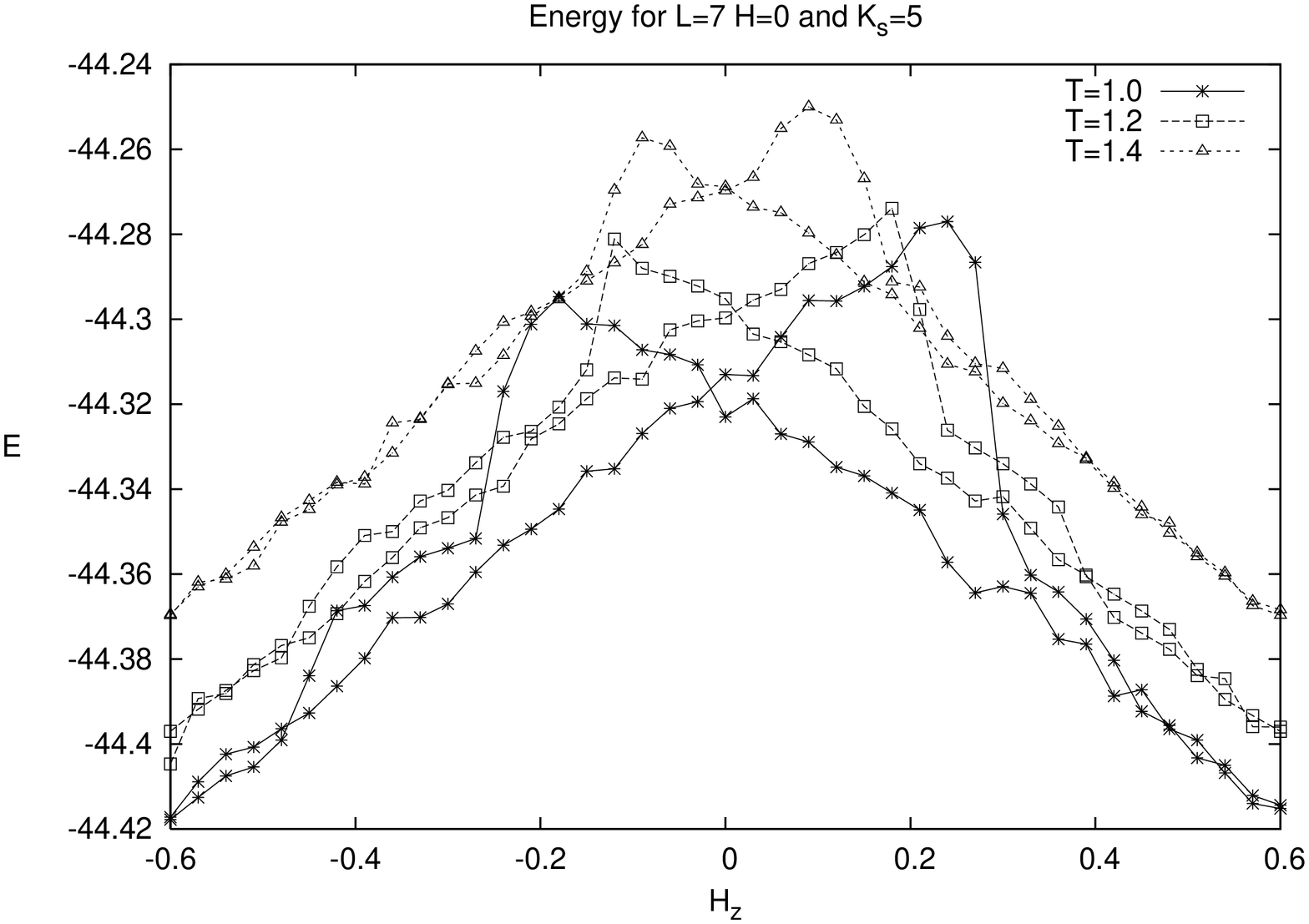}
\caption{Hysteresis loop for $L=7$ and $K_s=5$ in the zero field cooled case. The upper panel shows the $z$-component of the total
magnetization as a function of the applied field in the range $1 \le T \le 5$. The middle panel shows the behaviour over a narrower
temperature range. The lower panel shows the energy as a function of field for $T=1, 1.2, 1.4$.
}
 \label{fig8}
\end{figure}

Fig. \ref{fig8} shows typical hysteresis loops for $L=7$ and $K_s=5$ for the zero-field cooled case at various temperatures. The upper panel
corresponds to the range $1 <T <5$ and indicates that the width of the hysteresis loop quickly narrows as the blocking temperature is
reached.  $T_B$ was estimated to  be  $\sim 3$ for this size of particle. The middle panel shows hysteresis loops in the narrower
range $1 < T < 1.4$. As the applied field is reduced in magnitude and reversed, the component of the magnetization in the $z$-direction slowly
decreases until it rapidly changes direction near $H_z \sim -0.2$. However, the reversal is not complete until $H_z \sim -0.4$. Similar behaviour
is observed along the return path as $H_z$ is increased to positive values. At the lowest temperature $T=1$, the coercive field (as measured by
when the magnetization changes sign) is $H_c \sim 0.25$. The $T=1$ loop appears to have sub-loops when the magnitude of the applied field is between 0.2 and 0.4  which indicate that complete reversal does not occur immediately.

The lower panel in Fig.~\ref{fig8} shows the energy per particle as a function of the applied field for $T=1, 1.2, 1.4$. As the field is reduced from $H_z=0.6$, the energy increases reaching a maximum near
$H_z \sim -0.2$ and drops abruptly as the field is reduced further.
A second abrupt decrease occurs near $H_z \sim -0.4$. These
abrupt decreases indicate the limits of metastable configurations of the particle as the applied field is changed. As the temperature increases, the size of the
abrupt energy jumps decrease and their location moves towards $H_z=0$. Hence the reversal appears to be a two step process.

\begin{figure}[htb]
 \centering
\includegraphics[width=2.5in,angle=-90]{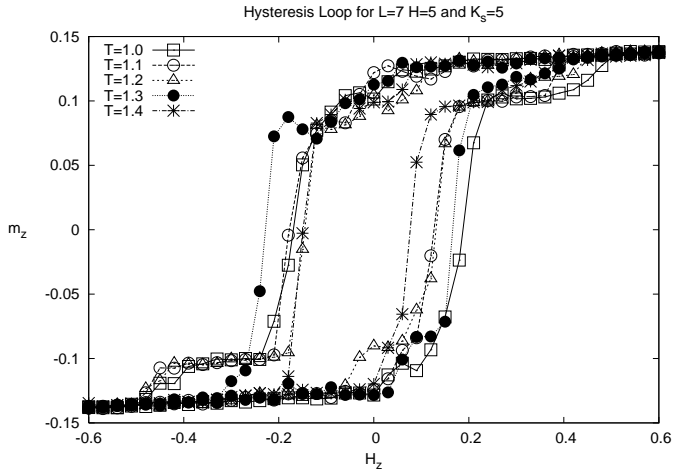}
\caption{Hysteresis loop for $L=7$ and $K_s=5$ in the non-zero field cooled case. 
}
 \label{fig9}
\end{figure}

Fig. \ref{fig9} shows the hysteresis loops for a particle which has been cooled in a field $H=5$ applied in the $z$-direction.
The behaviour is very similar to that in the zero-field cooled case shown in Fig. \ref{fig8} except that the hysteresis loop
appears to have a small shift to the left and is no longer symmetric about $H_z=0$. This feature indicates that a small amount of exchange bias may be present. However, this shift disappears as soon as the temperature is increased. Both
the core and surface exhibit the two stage reversal process for this value of $K_s$. This suggests that the particle reversal is quite different from a single domain
reversal mechanism where the core reversal occurs separately and is affected by the surface. On the contrary, different regions of the particle have their core
and surface acting together at two different reversed fields.

\begin{figure}[htbp]
 \centering
\includegraphics[width=2.5in,angle=-90]{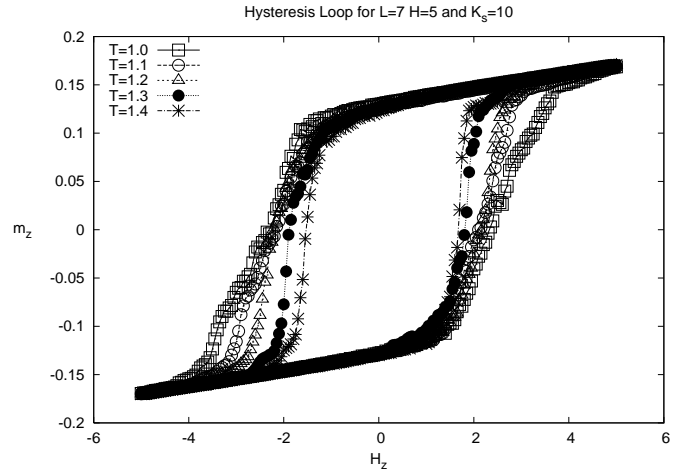}
\includegraphics[width=2.5in,angle=-90]{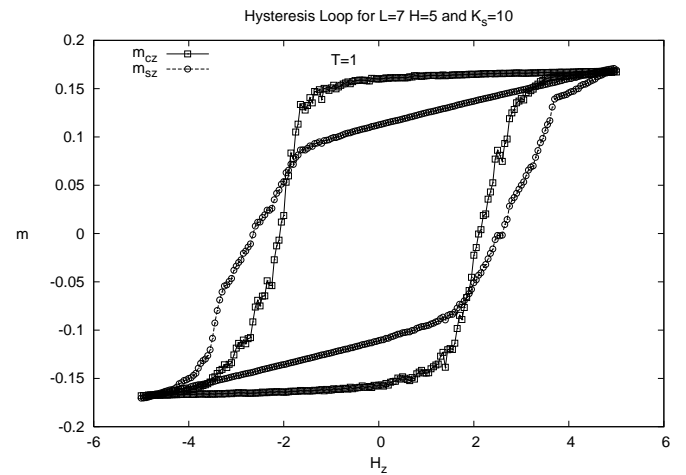}
\caption{Hysteresis loops for $L=7$ and $K_s=10$ at low temperatures (upper panel) in the field cooled case and the core and surface contributions at $T=1$ (lower panel).
at $T=1$.}
 \label{fig10}
\end{figure}

We have also studied the hysteresis properties for smaller and larger values of the surface anisotropy constant $K_s$.
Fig. \ref{fig10} shows results for $L=7$ and $K_s=10$ after cooling in a field. The upper panel shows $m_z$ at several low temperatures where a large
coercivity is present. The two step reversal process found for $K_s=5$ now seems to be replaced by a continuous
process which occurs over a wide range of reversal fields. The lower panel shows the hysteresis loop for both the core and surface magnetizations when $K_s=10$
at the lowest temperature $T=1$. The core and surface magnetizations reverse at different fields. As the applied field
is reduced from it's maximum value, the surface magnetization decreases linearly until the point where the core magnetization changes sign. This is followed by another linear region which has a much larger slope. Hence, for larger values of the surface anisotropy, the reversal of the particle's magnetization again appears to be a two stage process but it is quite different from that at smaller values of $K_s$.

\begin{figure*}[htbp]
 \centering
\includegraphics[width=1.5in,angle=-0]{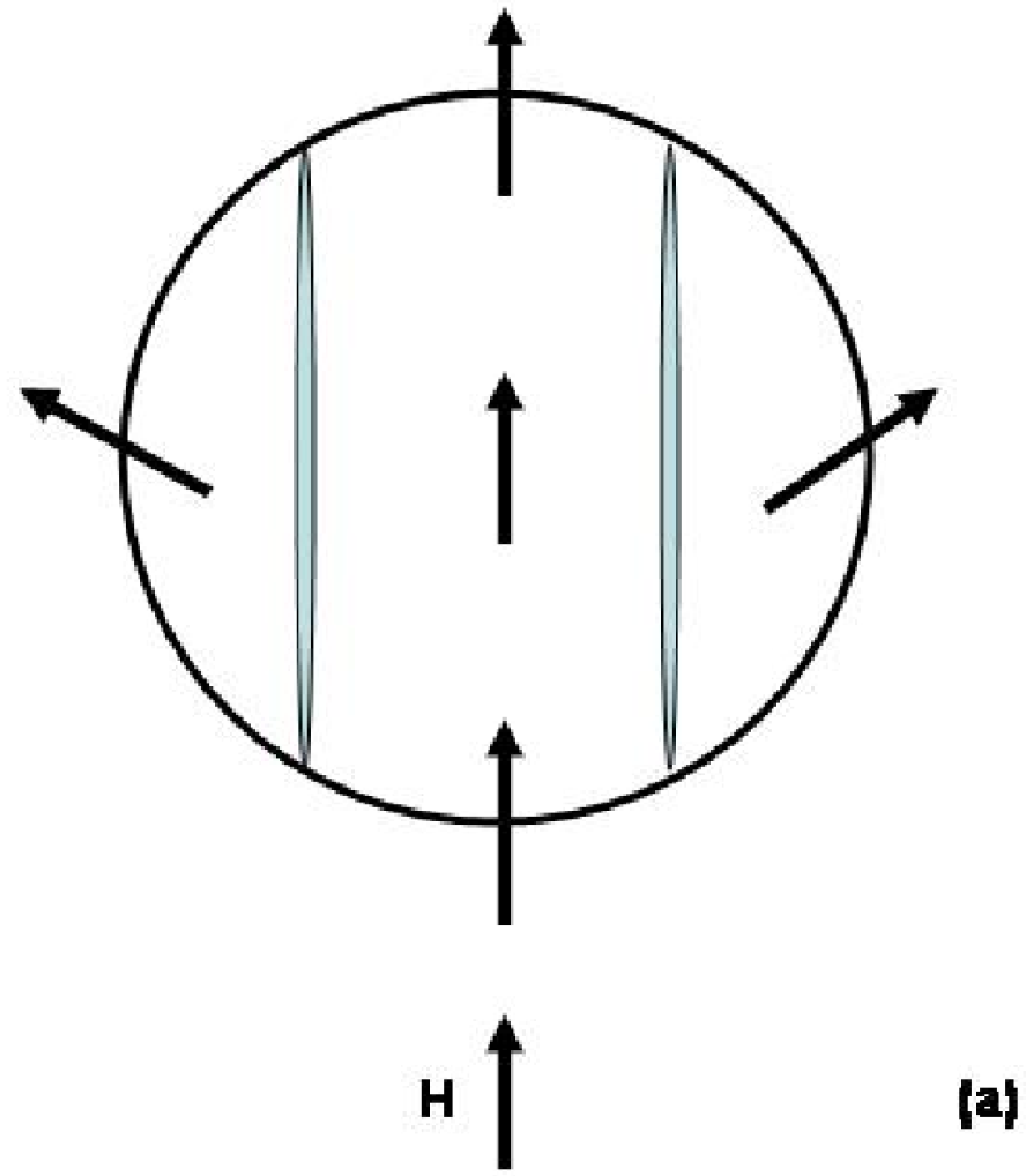}
\includegraphics[width=1.5in,angle=-0]{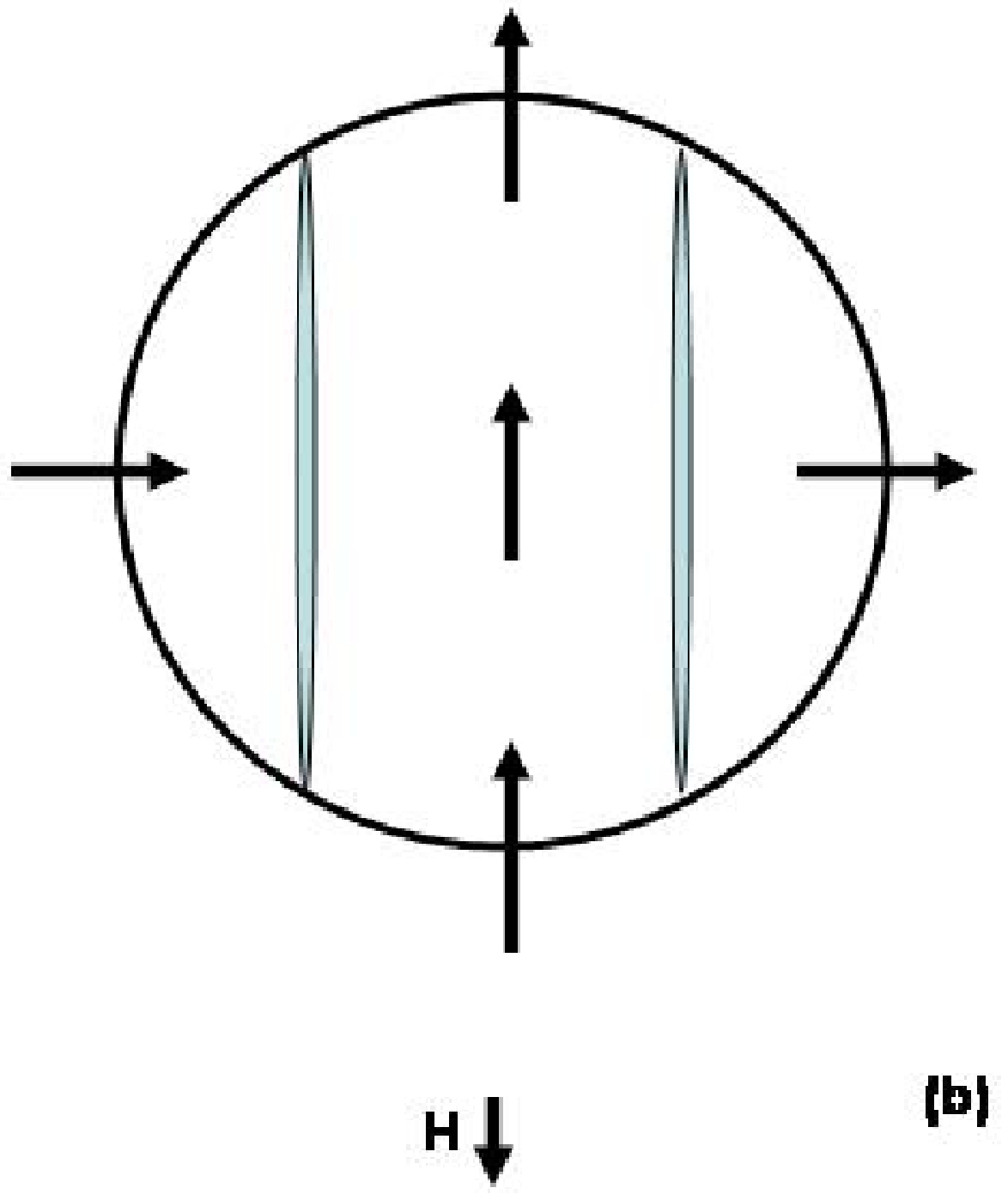}
\includegraphics[width=1.5in,angle=-0]{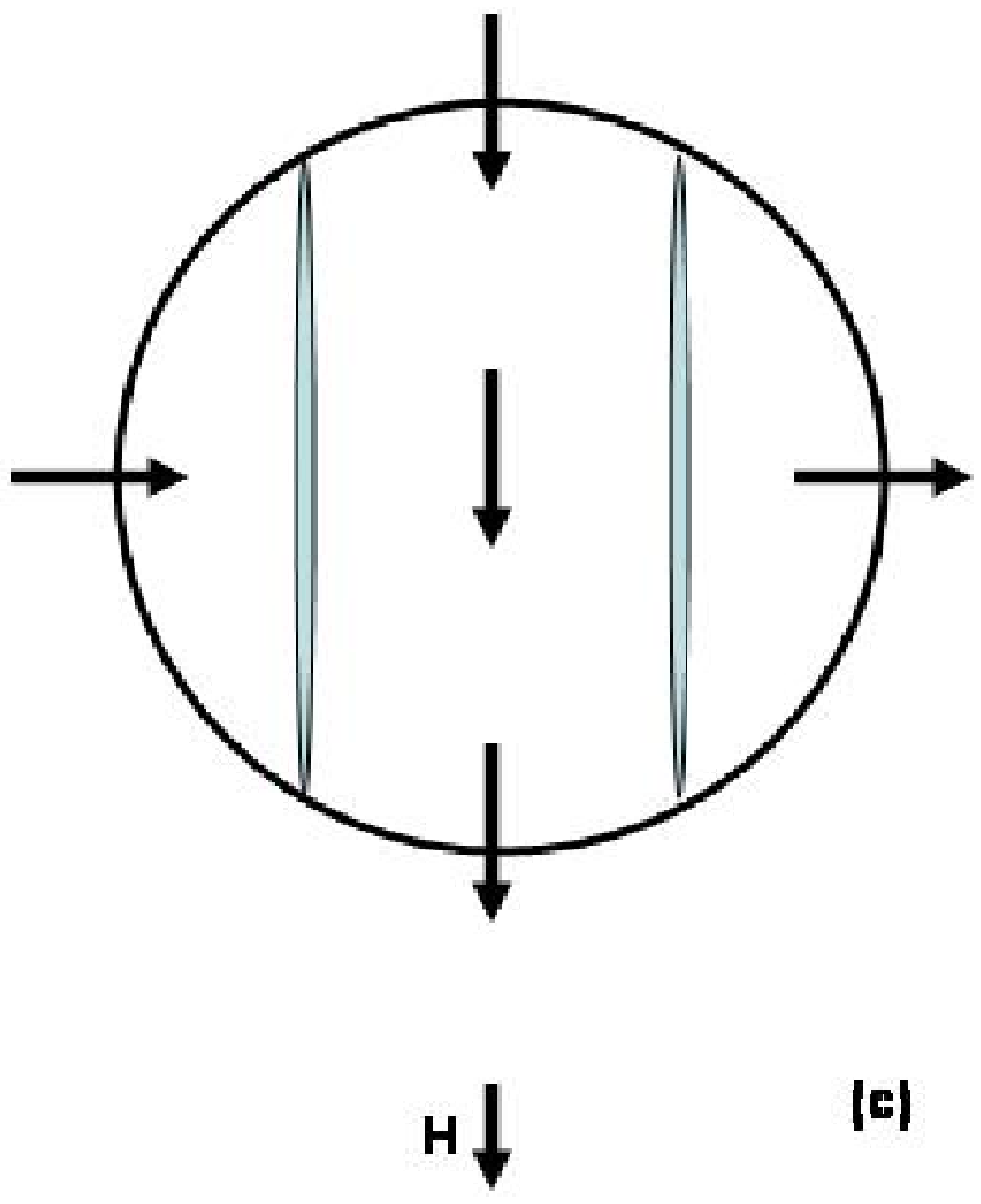}
\includegraphics[width=1.4in,angle=-0]{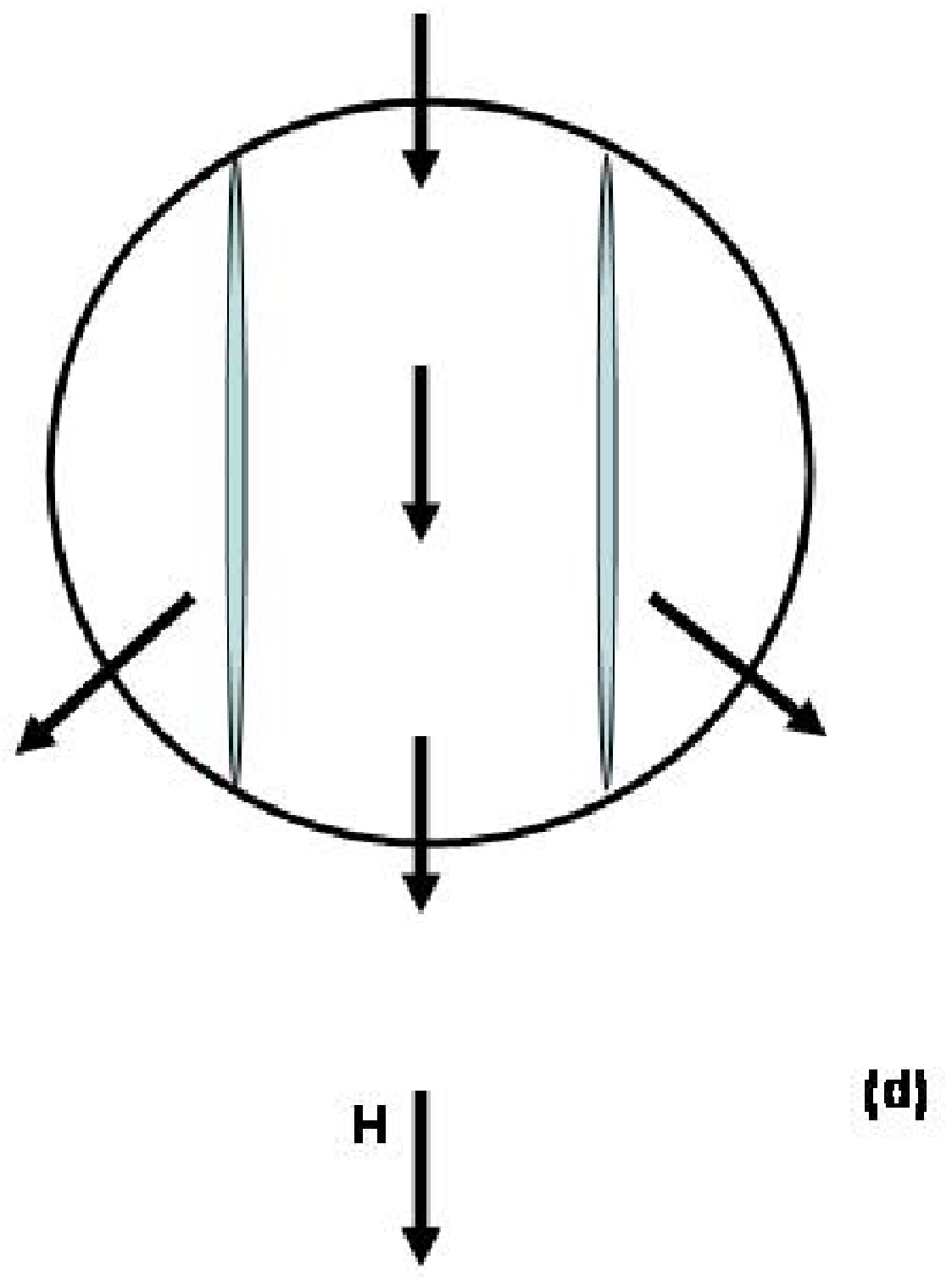}

\caption{Schematic description of the magnetization reversal process for $K_s=5$.
}
 \label{fig11}

\includegraphics[width=1.6in,angle=-0]{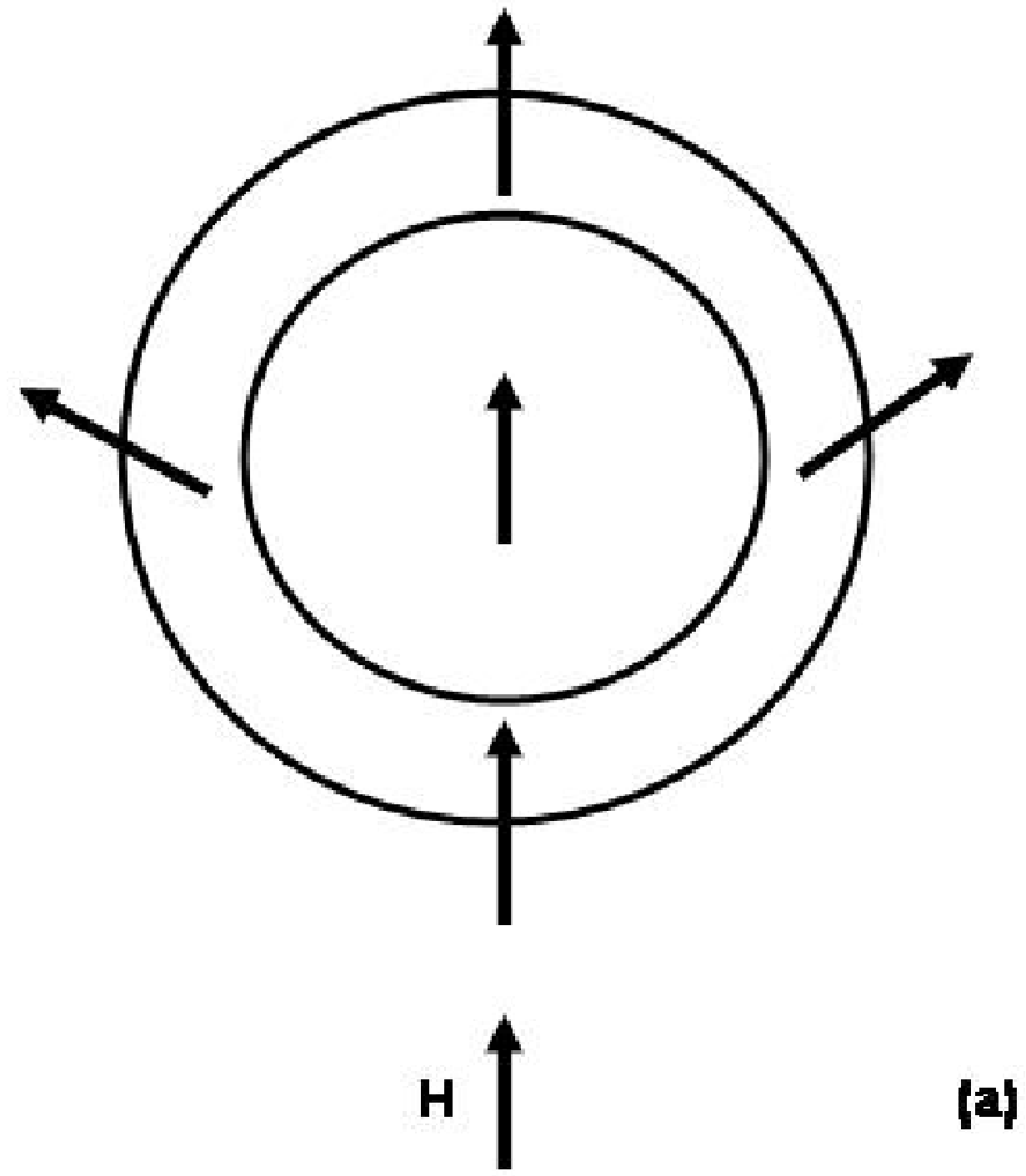}
\includegraphics[width=1.5in,angle=-0]{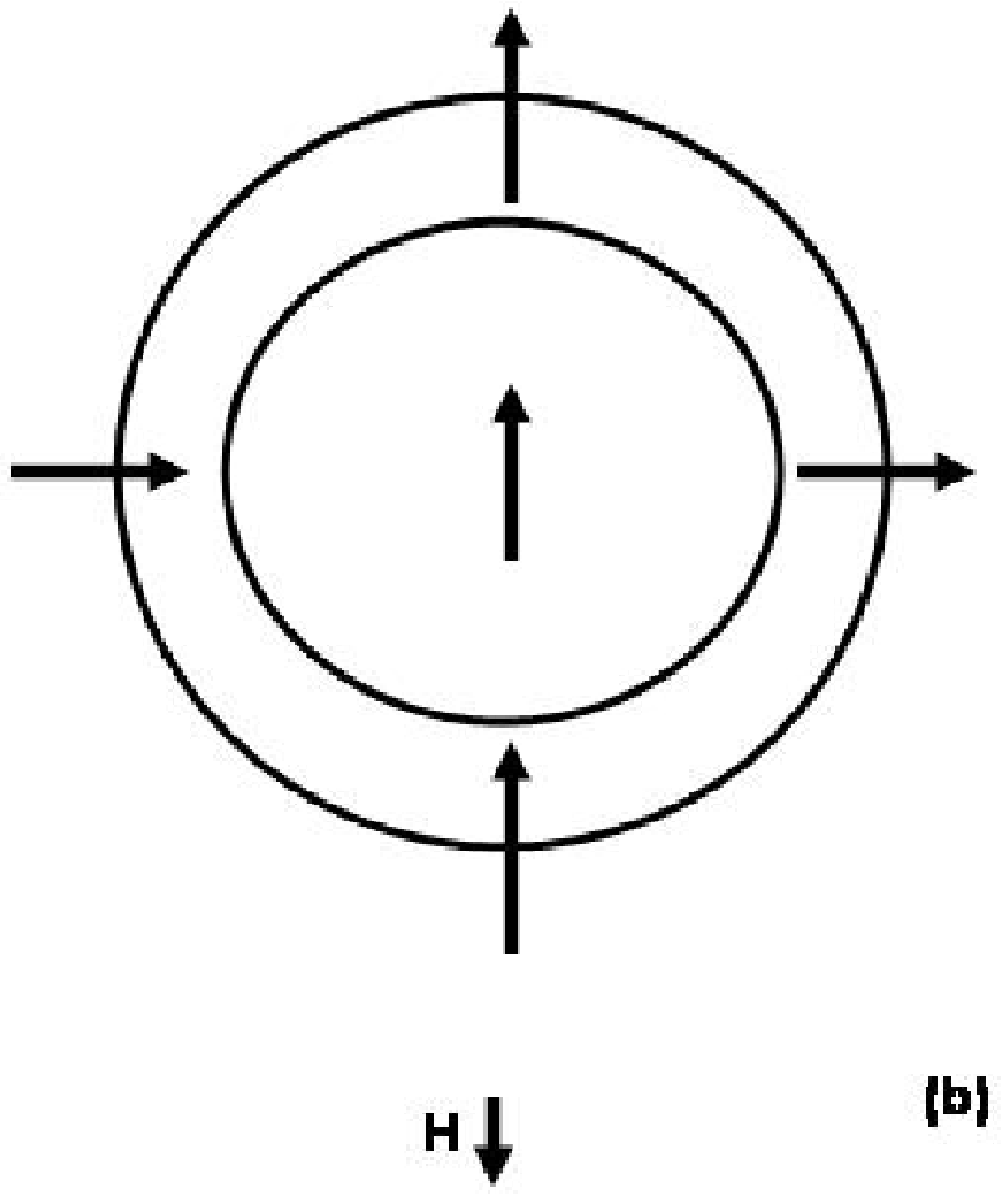}
\includegraphics[width=1.5in,angle=-0]{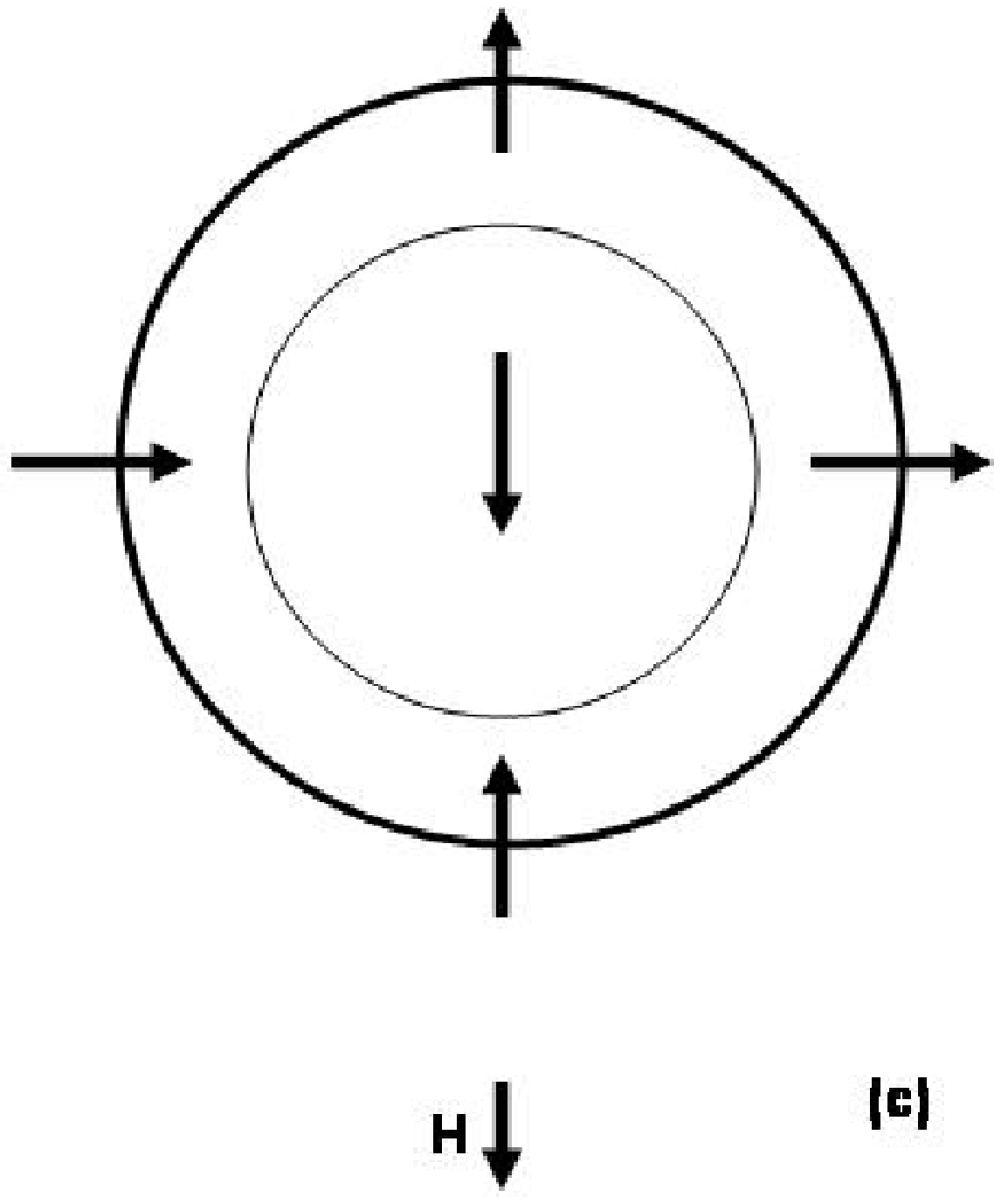}
\includegraphics[width=1.5in,angle=-0]{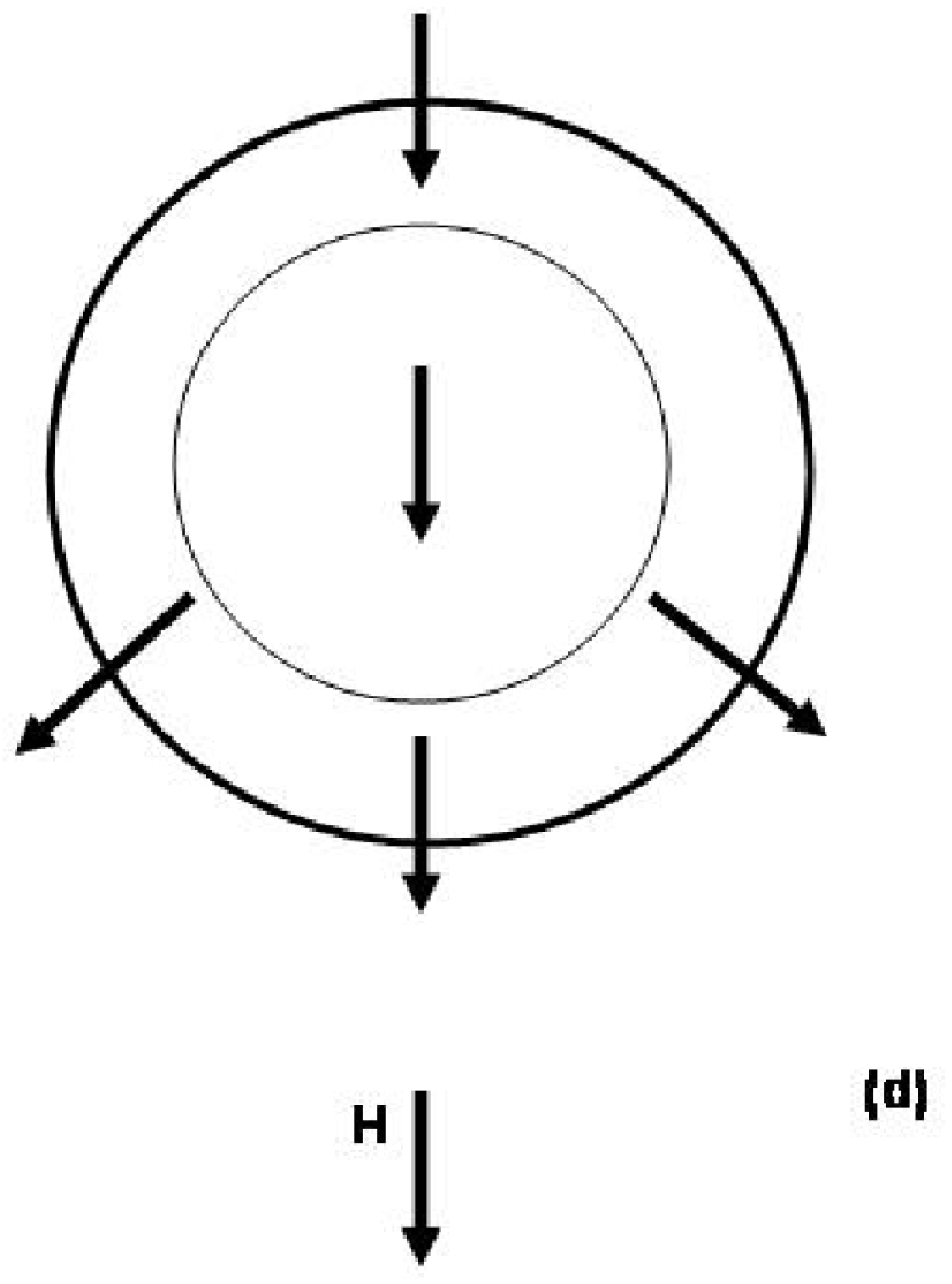}

\caption{Schematic description of the magnetization reversal process for $K_s=10$.
}

 \label{fig12}
\end{figure*}

Fig. \ref{fig11} illustrates schematically the two stage reversal process for the case $K_s=5$. In frame (a), the core and surface magnetizations are
polarized in the direction of the applied field. As the field is reduced and reversed as in frame (b), the magnetizations of both the core and surface in the
equatorial region of the particle develop components perpendicular to the field. Note that the figure is a cut through the particle in the $x-z$ plane. Since the anisotropy favors the moments to be normal to the
surface, it is this region which can overcome the applied field first. As the field is reversed further as in frame (c), the core and surface spins in the polar
regions reverse but the equatorial regions maintain a large transverse component. Finally in frame (d), the core and surface magnetizations are
polarized in the reverse direction. Similar reversal processes have been reported by Berger et. al. \cite{Tamin}

Fig. \ref{fig12} shows a similar picture for the case when $K_s=10$. Stages (a) and (d) are the same as in the previous case but stages (b)
and (c) are different. In (c), the core magnetization reverses but the surface magnetization does not. The surface magnetization reverses
when the applied field is reduced further as in (d). 

\section{Discussion and Conclusion}

In this paper we have used Monte Carlo methods to study the magnetic properties of maghemite nanoparticles.  
The Monte Carlo method employed
is a modified version of the usual heat bath method used to study classical spins to allow
for the finite length of the Fe spins in maghemite. The modifications only affect the behaviour of the thermodynamic quantities at low
temperatures where classical models fail.  At high temperatures the core spins begin to develop a ferrimagnetic order with a net moment. However, the direction of the moment is fluctuating and the particle exhibits superparamagnetic behaviour down to
much lower temperatures. The surface spins do not develop ferrimagnetic order until a much lower temperature $T_s \sim 5$. The net moment
of the particle continues to fluctuate until the blocking temperature $T_B \sim 3$ is reached and a coercivity develops. The temperature dependence of the surface magnetization obtained from our Monte Carlo results is very similar to that observed
experimentally\cite{Vanlierop}.

Both the surface ordering temperature and the blocking temperature do not depend on the size of the nanoparticle. However, they both increase
approximately linearly with the surface anisotropy constant $K_s$. The fact that $T_B$ does not depend on particle size in the range of sizes studied indicates that the effective anisotropy barrier is temperature dependent and leads to an effective anisotropy $K_{eff} \sim 1/V$. Increases of $K_{eff}$ with decreasing nanoparticle size have
been observed experimentally\cite{bodker,skoropata,he}.

  When the particles are cooled in an applied field, the magnetizations
increase. The largest change occurs for the surface magnetization.
Hysteresis loops were studied in both the zero field and non-zero field cooled cases. The differences in the loops for these different cooling
histories were quite small. Experimental systems exhibit a shift of the hysteresis loop when cooled in a field. This shift corresponds to an exchange bias
field. We find little evidence of exchange bias in our model except at the lowest temperature.

The surface anisotropy has a strong effect on the hysteresis loops and the nature of the moment reversal of the particles. For small
values of $K_s$, the blocking temperature is reduced significantly and the hysteresis loops only display superparamagnetic behaviour
with no coercivity. The particles reverse their magnetization by following the field direction. For intermediate values of $K_s$, the blocking
temperature is increased and the hysteresis loops exhibit coercivity. There is a main loop as well as minor loops. The reversal of the
moment appears to be a two stage process. At large values of the surface anisotropy, the blocking temperature is increased again
and the hysteresis loops exhibit a ten-fold increase in the coercivity. The minor loops are merged with the main loop and the moment reversal
occurs in two steps with the core spins reversing first followed by the surface spins.

The results presented here\cite{adebayo} are our first attempt at studying the magnetic properties of nanoparticles. Some of the results agree
qualitatively with recent experiments but others do not. The absence of exchange bias in our model could be due to the fact that
our model is too simple or it could be due to our Monte Carlo algorithm. The Monte Carlo algorithm is a method used to study equilibrium properties of magnetic models, whereas, hysteresis and exchange bias are determined by metastable configurations. Further work should include modifying the model or the algorithm or both.
One of the problems with our calculations is that they require a great deal of cpu time. Each hysteresis loop uses 30,000 mcs for
each of 200 field values at a single temperature which typically requires 24 hours to complete. Parallelization of the code could
speed up this process significantly and allow us to vary both the exchange parameters and applied field strengths over a wider range. A study of the same nanoparticles arranged on a lattice would require such improvements.

\begin{acknowledgments}
This work was supported by the Natural Sciences and Research Council of Canada. We thank J. van Lierop for many
useful discussions on nanomagnetism.
\end{acknowledgments}
\bibliographystyle{unsrt} 
\bibliography{nano}

\begin{thebibliography}{10}

\bibitem{kas}
O.~Kasyutich, R.~D. Desautels, B.~W. Southern, and J.~van Lierop.
\newblock Novel aspects of magnetic interactions in a macroscopic 3d
  nanoparticle-based crystal.
\newblock {\em Phys. Rev. Lett. (to appear)}, 2010.

\bibitem{Plumer.2009}
M.~L. Plumer, J.~van Lierop, B.~W. Southern, and J.~P. Whitehead.
\newblock Micromagnetic simulations of interacting dipoles on a fcc lattice:
  Application to nanoparticle assemblies.
\newblock {\em submitted}, 2010.

\bibitem{Coey}
J.~M.~D. Coey.
\newblock {\em Phys. Rev. Lett.}, {\bf 27}:1140, 1971.

\bibitem{Millan}
A.~Millan, A.Urtizberea, N.J.O. Silva, F.Palacio, V.S. Amaral, E.~Snoeck, and
  V.~Serin.
\newblock {\em J.Magn.Magn.Mater.}, {\bf 312}:L5, 2007.

\bibitem{Vanlierop}
T.N. Shendruk, R.D. Desautels, B.W. Southern, and J.~van Lierop.
\newblock {\em Nanotechnology}, {\bf 18}:455704, 2007.

\bibitem{Brown}
William Fuller~Brown Jr. and Clarke E.~Johnson Jr.
\newblock {\em J. Appl. Phys.}, {\bf 33}:2752, 1962.

\bibitem{Costa}
G.M. da~Costa, E.~de~Grave, and R.E Vandenberghe.
\newblock {\em Hyperfine Interact.}, {\bf 117}:207, 1998.

\bibitem{Dimi2}
Radek Zboril, Miroslav Mashlan, and Dimitris Petridis.
\newblock {\em Chem.Mater.}, {\bf 14}:969, 2002.

\bibitem{ZBoril}
J.~T\H{u}cek and R.~Zboril.
\newblock {\em Czechoslovak Journal of Physics}, {\bf 55}:893, 2005.

\bibitem{Dimi1}
J.~T\H{u}cek, Radek Zboril, and Dimitris Petridis.
\newblock {\em J. Nanosci. Nanotechnol.}, {\bf 6}:926, 2006.

\bibitem{Berk}
R.H. Kodama and A.E Berkowitz.
\newblock {\em Phys. Rev. B.}, {\bf 59}:6321, 1999.

\bibitem{Kodama}
R.H. Kodama, A.E. Berkowitz, E.J. McNiff, and S.~Foner.
\newblock {\em Phys. Rev. Lett.}, {\bf 77}:394, 1996.

\bibitem{Linder}
S.M{\o}rup, F.~B{\o}dker, P.V. Hendriksen, and S.~Linderoth.
\newblock {\em Phys. Rev. B.}, {\bf 52}:287, 1995.

\bibitem{Labaye1}
J.~Restrepo, Y.~labaye, and J.M. Greneche.
\newblock {\em Revista Colombiana De Fisica}, {\bf 38}:1559, 2006.

\bibitem{Labaye2}
J.~Restrepo, Y.~labaye, and J.M. Greneche.
\newblock {\em Revista Colombiana De Fisica}, {\bf 39}:183, 2007.

\bibitem{Labaye3}
J.~Restrepo, Y.~labaye, and L.~Berger~J.M. Greneche.
\newblock {\em J. Magn.Magn.Mater.}, {\bf 272-276}:681, 2004.

\bibitem{Labaye4}
J.~Restrepo, Y.~labaye, and J.M. Greneche.
\newblock {\em Physica B}, {\bf 384}:221, 2006.

\bibitem{metrop}
M.~Metropolis, A.~W. Rosenbluth, M.~N. Rosenbluth, A.~H. Teller, and E.~Teller.
\newblock {\em J. Chem. Phys.}, 21:1087, 1953.

\bibitem{Eftax}
E.~Eftaxias and K.N. Trohidou.
\newblock {\em Phys. Rev. B}, {\bf 71}:134406, 2005.

\bibitem{Troh}
K.N. Trohidou, M.~Vasilakaki, L.~Del Bianco, D.~Fiorani, and A.M. Testa.
\newblock {\em J. Magn. Magn. Mater.}, {\bf 192}:203, 1999.

\bibitem{Vasi}
Marianna Vasilakaki and Kalliopi~N Trohidou.
\newblock {\em J.Phys.D:Appl.Phys.}, {\bf 41}:134006, 2008.

\bibitem{Oscar1}
$\grave{O}$scar Iglesias, Xavier Batlle, and Am$\acute{i}$car Labarta.
\newblock {\em J.Phys.D:Appl.Phys.}, {\bf 41}:134010, 2008.

\bibitem{Oscar2}
$\grave{O}$scar Iglesias, Xavier Batlle, and Am$\acute{i}$car Labarta.
\newblock {\em Phys. Rev. B}, {\bf 72}:212401, 2005.

\bibitem{Amilcar}
$\grave{O}$scar Iglesias and Am$\acute{i}$car Labarta.
\newblock {\em Phys. Rev. B}, {\bf 63}:184416, 2008.

\bibitem{Igle}
$\grave{O}$scar Iglesias and Am$\acute{i}$car Labarta.
\newblock {\em Phys. stat.sol.}, {\bf 1}:3481, 2004.

\bibitem{Kach}
H.~KachKachi, A.~Ezzir, M.Nogues, and E.Tronc.
\newblock {\em Eur. Phys. J. B}, {\bf 14}:681, 2000.

\bibitem{biasi}
E.~De Biasi, C.A. Ramos, R.D. Zysler, and D.~Fiorani.
\newblock {\em Physica B}, {\bf 372}:345, 2006.

\bibitem{Mazo}
J.~Mazo-Zuluaga, J.~Restrepo, and J.~Mej$\acute{i}$a-L$\acute{o}$pez.
\newblock {\em J. Appl. Phys.}, {\bf 103}:113906, 2008.

\bibitem{Mazo1}
J.~Mazo-Zuluaga, J.~Restrepo, and J.~Mej$\acute{i}$a-L$\acute{o}$pez.
\newblock {\em Physica B}, {\bf 398}:187, 2007.

\bibitem{Mazo2}
J.~Mazo-Zuluaga, J.~Restrepo, and J.~Mej$\acute{i}$a-L$\acute{o}$pez.
\newblock {\em J. Phys.:Condens.Matter}, {\bf 20}:195213, 2008.

\bibitem{Usov}
N.A. Usov and Yu.~B. Grebenshchikov.
\newblock {\em J. Appl. Phys.}, {\bf 104}:043903, 2008.

\bibitem{Miyatake}
Y~Miyatake, M~Yamamoto, J.J. Kim, M.~Toyonaga, and O.Nagai.
\newblock {\em J.Phys.C:Solid State Phys.}, {\bf 19}:2539, 1986.

\bibitem{Lee}
L.~W. Lee and A.P. Young.
\newblock {\em Phys. Rev. B}, {\bf 76}:024405, 2007.

\bibitem{neel}
L.~N$\acute{e}$el.
\newblock {\em Ann. Geophys.}, {\bf 5}:99, 1949.

\bibitem{brown5}
W.F. Brown.
\newblock {\em Phys. Rev.}, {\bf 130}:1677, 1963.

\bibitem{bodker}
F.~B$\o$dker, S.~M$\o$rup, and S.~Linderoth.
\newblock {\em Phys. Rev. Lett.}, {\bf 72}:282, 1994.

\bibitem{skoropata}
E.~Skoropata, R.D. Desautels, and J.~van Lierop.
\newblock {\em J. Appl. Phys.}, {\bf 105}:07B503, 2009.

\bibitem{he}
Lin He.
\newblock {\em Solid State Commun.}, {\bf 150}:743, 2010.

\bibitem{Tamin}
L.~Berger, Y.~Labaye, M.~Tamine, and J.~M.~D. Coey.
\newblock {\em Phys. Rev. B}, {\bf 77}:104431, 2008.

\bibitem{adebayo}
Kenneth Adebayo.
\newblock Master's thesis, University of Manitoba, Winnipeg Manitoba, Canada,
  2009.

\end{thebibliography}
\end{document}